\documentclass [12pt]{article}
\RequirePackage[english]{babel}
\RequirePackage[cp1251]{inputenc}
\usepackage{amsmath,amssymb}

\begin{document}
\sloppy
\begin{center}

{\Large \bf F.A. Gareev,  I.E. Zhidkova}\\ \vspace*{0.5cm} {\large
\bf Quantization of Differences Between Atomic and Nuclear Rest
Masses and Self-organization of Atoms and Nuclei}\\
\vspace*{1cm}
{\sl Joint Institute for Nuclear Research, Dubna, Russia\\
 e-mail:gareev@thsun1.jinr.ru}\\
\end{center}

\begin{abstract}
 We come to the conclusion that  all atomic models
based on either  the Newton equation and the Kepler laws, or  the
Maxwell equations, or  the Schrodinger and Dirac equations are in
reasonable agreement with experimental data. We can only suspect
that these equations are grounded on the same fundamental
principle(s) which is (are) not known or these equations can be
transformed into each other. We proposed a new mechanism of LENR:
cooperative processes in the whole system - nuclei+atoms+condensed
matter - nuclear reactions in plasma - can occur at smaller
threshold energies than the  corresponding ones on free
constituents. We were able to quantize  phenomenologically the
first time the differences between atomic and nuclear rest masses
by the formula (in MeV/$c^{2}$) $\Delta
M=\frac{n_{1}}{n_{2}}*0.0076294,\; n_{i}=1,2,3,...$ Note that this
quantization rule is justified for atoms and nuclei with different
$A,\;N$ and $Z$ and the nuclei and atoms represent a coherent
synchronized systems - a complex of coupled oscillators
(resonators). The cooperative resonance synchronization mechanisms
are responsible for explanation of how the electron volt world can
influence  the nuclear mega electron volt world. It means that we
created new possibilities for inducing and controlling nuclear
reactions by atomic processes.
\end{abstract}
\section{Introduction}
 The  review of possible stimulation mechanisms of LENR (low energy nuclear reaction) is presented in \cite{2}.
We have concluded that transmutation of nuclei at low energies and
excess heat are possible in the framework of the known fundamental
physical laws – the universal cooperative resonance
synchronization principle \cite{1}, and different enhancement
mechanisms of reaction processes \cite{2}. The superlow energy of
external fields, the excitation and ionization of atoms may play
the role of a trigger for LENR. Superlow energy of external fields
may stimulate LENR \cite{3}. We bring  strong arguments that the
cooperative resonance synchronization mechanisms are responsible
for explanation of how the electron volt world can influence  the
nuclear mega electron volt world \cite{3}.

Nuclear physicists are absolutely sure that this  cannot happen.
Almost all nuclear experiments were carried out in conditions when
colliding particles interacted with the nuclear targets which
represented a gas or a solid body. The nuclei of the target are in
the neutral atoms surrounded by orbital electrons. All existing
experimental data under such conditions teach us that nuclear low
energy transmutations  are not observed due to the Coulomb
barrier.

LENR with transmutation of nuclei occurs in different conditions
and different processes (see, for example, publications in
http://wwww.lenr-canr.org/ which contain more than 500 papers) but
these processes have  common properties: interacting nuclei are in
the ionized atoms or completely without electrons (bare nuclei).
Therefore, LENR with bare nuclei and nuclei in ionized atoms
demonstrated a drastically different properties in comparison with
nuclei in neutral atoms \cite{1}. This is nuclear physics in
condensed matter or in plasma state of matter. For example, the
measured half-life \cite{3c} for bare $^{187}Re^{75+}$ of
$T_{1/2}=(32.9\pm2.0)$ yr is billion times shorter than that for
neutral $^{187}Re$.

Natural geo-transmutations of nuclei in the atmosphere and earth
are established very well \cite{3d,3g,3f}. They occur at the
points of a strong change in geo- and electromagnetic fields.
Moreover, there are a lot of experimental data \cite{3v} that the
nuclear fusion and transmutation in biological systems are the
real phenomena.

 Nucleons in nuclei and electrons in atoms represent the whole
system in which all motions are synchronized and self-sustained
(see our publications \cite{1,2,3,3a} and Shadrin V.N. talk
\cite{3b}). Nucleons in nuclei and electrons in atoms are
nondecomposable into independent motions of nucleons and
electrons.

Investigation of this phenomenon requires the knowledge of
different branches of science: nuclear and atomic physics,
chemistry and electrochemistry, condensed matter and solid state
physics, astrophysics, biology, medicine, ...

$\bullet$The differentiation of science which was rather useful at
the beginning brings civilization to a catastrophe.  Therefore the
integration of different branches of science is a question of
vital importance.

 The puzzle of poor reproducibility of experimental data is the fact that LENR occurs in open systems and
it is extremely sensitive to parameters of external fields and
systems. The classical reproducibility principle should be
reconsidered for LENR experiments. Poor reproducibility and
unexplained results do not mean the experiment is wrong.  Our main
conclusion is:
 LENR may be understood in terms of the known fundamental laws without any violation of  basic physics.
The fundamental laws of physics should be  the same in micro- and
macrosystems.

$\bullet$ LENRs take places in open systems in which all
frequencies and phases are coordinated according to the universal
cooperative resonance synchronization principle. Poor
reproducibility of experimental results and extreme difficulties
of their interpretation in the framework of modern standard
theoretical physics (there are about 150 theoretical models
\cite{3t} which are not accepted by physical society) are the main
reasons for the persistent nonrecognition of cold fusion and
transmutation phenomenon.

Recent  progress in both directions is remarkable (see
http://www.lenr-canr.org/, http://www.iccf12.org/,
http://www.iscmns.org/);in spite of being rejected by physical
society, this phenomenon is a key point for further success in the
corresponding fundamental and applied research. The results of
this research field can provide  new ecologically pure sources of
energy, substances, and technologies.

The possibilities of inducing and controlling nuclear reactions at
low temperatures and pressures by using different low-energy
fields and various physical and chemical processes were discussed
in \cite{2,3,3a}. The aim of this paper is to present the results
of phenomenological quantization of nuclear and atomic mass
differences which can bring new possibilities for inducing and
controlling nuclear reactions by atomic processes and new
interpretation of self-organizations of the hierarchial systems in
the Universe including the living cells. How do the atoms and
nuclei have their perpetual motions? How is the Universe
constructed? How it links the smallest structures in the Universe
to the largest? The three questions are interconnected.

Let us start with the description of  the hydrogen atom structure
in different models using the standard basic physics that is well
established, both theoretically and experimentally in micro- and
macrosystems.

\section{The Bohr Model}

At the end of the 19th century it was established that the
radiation from hydrogen was emitted at specific quantized
frequencies. Niels Bohr developed the model to explain this
radiation using four postulates:

1. An electron in an atom moves in a circular orbit about the nucleus under the influence of the
Coulomb attraction between the electron and the nucleus, obeying the laws of classical mechanics.

2. Instead of the infinity of orbits which would be possible in
classical mechanics, it is only possible for an electron to move
in an orbit for which its orbital angular momentum $L$ is
integral multiple of $\hbar$:
                                    $$ L=n\hbar,\;\;n=1,2,3,…\eqno(1) $$

3. Despite the fact that it is constantly accelerating, an
electron moving in  such an allowed orbit does not radiate
electromagnetic energy. Thus, its total energy $E$ remains
constant.

4. Electromagnetic radiation is emitted if an electron, initially
moving in an orbit of total energy $E_{i}$, discontinuously
changes its motion so that it moves in an orbit of total energy
$E_{f}$. The frequency $\nu$ of the emitted radiation is equal to
the quantity

$$\nu_{if}=\frac{E_{i}-E_{f}}{h},\eqno(2)$$

where $h$ is Planck's constant.
 The electron is held on a stable
circular orbit around the proton. The hydrogen atom consists of
one heavy proton in the center of atom with one lighter electron
orbiting around proton. The Coulomb force is equal to the
centripetal force, according to Newton's second law

$$\frac{e^{2}}{r^{2}}=\frac{mv^{2}}{r},\eqno(3)$$
where $r$ is is the radius of the electron orbit, and $v$ is the
electron speed. The force is central; hence from the quantization
condition (1) we have

$$L=\mid \vec{r}*\vec{p}\mid=mvr=n\hbar.\eqno(4)$$

After solving equations (3) and (4) we have

$$v=\frac{e^{2}}{n\hbar},\;r=\frac{n^{2}\hbar^{2}}{me^{2}}=n^{2}a_{0}.\eqno(5)$$
Following equation (3) the kinetic energy is equal to

$$E_{k}=\frac{1}{2}mv^{2}=\frac{e^{2}}{2r},\eqno(6)$$
and hence the total energy is

$$E=E_{k}+V=\frac{e^{2}}{2r}-\frac{e^{2}}{r}=-\frac{e^{2}}{2r}.\eqno(7)$$

Having $r$ from equation (5) one can write the expression for the
energy levels for  hydrogen atoms

$$E=-\frac{me^{4}}{2\hbar^{2}n^{2}};\eqno(8)$$
the same results were further obtained  by quantum mechanics.

Using the angular momentum quantization condition $L=pr=nh/2\pi$
and Louis de Broglie's relationship $p=h/\lambda$  between
momentum and wavelength one can get

$$2\pi r=n\lambda. \eqno(9)$$

{\sl $\otimes$ It means that the circular Bohr orbit is an
integral number of the de Broglie wavelengths. }

\subsection{The Bohr 3th postulate}

Let us remember the Bohr 3th postulate: \\
$\bullet$ {\sl Despite the fact that it is constantly
accelerating, an electron moving in such an allowed orbit does not
radiate electromagnetic energy. Thus, its total energy $E$ remains
constant.}

The classical electrodynamics law:  an accelerating electron
radiates electromagnetic energies in full agreement with
experimental data.

Therefore, Bohr's postulate asserts that the classical
electrodynamics does not work on an atomic scale. Surprisingly,
the physical society accepts this postulate which declares that
the physical laws in macro- and microworld are different. An
electron in the Bohr model rotates around the motionless proton so
the proton is a nonactive partner. The motion of proton in
hydrogen atom was ignored completely -- fatal error of the Bohr
model.

We are convinced that the physical laws are unique and are the
same in different scale systems. The proton and an electron
represent two components of the same system --  a hydrogen atom.
We consider the hydrogen atom as the whole nondecomposable system
in which the motions of proton and electron are synchronized.
Therefore, if proton stops to move,  it should not be  allowed for
an electron to keep its own motion.

$\bullet$ {\sl The hydrogen atom in the ground state does not
radiate electromagnetic energy -- experimental fact.}

It was possible to describe the nonradiation hydrogen atom in the
ground state  even in 1913 on the basis of classical
electrodynamics as the result of standing wave formation in which
the motions of proton and electron are synchronized in such a way
that the electromagnetic energy flows are equal to zero.

Bohr's 3rd postulate can be reformulated in the following way: \\
$\bullet$ {\sl The hydrogen atom is an open system in which all
frequencies and phases of proton and electron are coordinated
according to the universal cooperative resonance synchronization
principle.}

 The proton and electron in a hydrogen atom form standing
electromagnetic waves so that the sum of radiated and absorbed
electromagnetic energy flows by electron and proton  is equal to
zero at distances larger than the orbit of electron \cite{5} – the
secret of success of the Bohr model (nonradiation of  the electron
on the stable orbit). The formation standing electromagnetic waves
with zero energy flows is the main reason of the hydrogen atom
stability.

\subsection{ Bohr's 4th Postulate}

Let us remember the Bohr 4th postulate:\\
$\bullet$ {\bf Electromagnetic radiation is emitted if an
electron, initially moving in an orbit of total energy $E_{i}$,
discontinuously changes its motion so that it moves in an orbit of
total energy $E_{f}$. The frequency $\nu$ of the emitted radiation
is equal to the quantity

$$\nu_{if}=\frac{E_{i}-E_{f}}{h},\eqno(2)$$

where $h$ is Planck's constant.}

The following simple question is: what is a source of information
that an electron knows in advance the value of emitted (absorbed)
energy. The answer is very simple.

We consider an electron motion in the one-dimensional infinite
potential well whose coordinates are equal to $z=-L/2$ and
$z=L/2$. Assume that an electron state is a superposition of the
ground state and first excited one
$$\psi(z,t)=\psi_{1}(z,t)+\psi_{2}(z,t),$$
$$\psi_{1}(z,t)=A_{1}e^{-i\omega_{1}t}cosk_{1}z,\;k_{1}L=\pi,$$
$$\psi_{2}(z,t)=A_{2}e^{-i\omega_{2}t}cosk_{2}z,\;k_{2}L=2\pi.$$
It is easy to calculate \cite{KRA64} the average value of a
$\overline{z}$ -- coordinate electron in one dimension potential:
$$\overline{z}=\frac{32L}{9\pi^{2}}\frac{A_{1}A_{2}}{A_{1}^{2}+A_{2}^{2}}cos(\omega_{2}-\omega_{1})t.$$

Therefore, the average position of a charge oscillates with the
frequency of beating $\omega_{beating}=\omega_{2}-\omega_{1}$.

$\bullet$ {\bf It means that the radiation frequency of the
electromagnetic waves is equal to the beating frequency between
the first excited state and the ground state
$$\omega_{radiation}=\omega_{beating}=\omega_{2}-\omega_{1},$$
according to the classical electrodynamics. An electron in the
mixed states knows the value of emitted (absorbed) energy in
advance which is equal to the beating frequency multiplied by
Planck's constant. This is the resonance process in which the de
Broglie wavelength $\lambda$ changes an integer number. Therefore,
the de Broglie wavelength plays the  role of the standard one. For
example, the musical instruments emit the sounds with frequencies
that are equal to instruments eigenfrequencies.}

The Bohr postulates were completely arbitrary and even violated
the well established laws of the classical electrodynamics.

 The standard point of view is that the Bohr model is actually accurate only for a
one-electron system, see below.

\subsection{The Sukhorukov Model -- Generalization\\ of the
Bohr-Sommerfield Model}

The Sukhorukov model \cite{SUK98}  is the generalization of the
Bohr-Sommerfield model for multi-electron atoms. Atoms have a
planetary structure. Rydberg's constant $R_{\inf}$ is the same for
all atoms. The ionization potentials have been calculated
\cite{SUK01} for 36 chemical elements with accuracy better than 1
eV.

\subsection{The Parson Model}

A.L. Parson \cite{PAR15} developed a model of  atoms in which each
electron forms a small magnet (in 1915). The rings of the charge
represent the shape of a toroid surrounding the nucleus. This
model was not accepted by physical society and was forgotten
despite that H.Stanley Allen \cite{ALL19} proved many outstanding
properties in comparison  with other models of atom.

\subsection{The Lucas Model of Atoms and Nuclei}

We quote D.L. Bergman's conclusion \cite{BER05}: {\sl In 1996,
while still a student in secondary school, Joseph Lucas introduced
his model of atom \cite{LUC96}. In this model, electrons, protons
and neutrons are all based on Bergman's Spinning Charge Ring Model
of Elementary Particles \cite{BER90,BER91} (a refinement of
Parson's Magneton). In terms of its predictive ability and
conformance with all known experiments, the Lucas Model of the
Atom is by far the most successful of all models of the atom ever
proposed. It is a physical model that shows where particles are
located throughout the volume of the atoms. This model predicts
the "magic numbers" 2,8,18, and 32 of electrons in the filled
shells and also able to predict why the Periodic Table of the
Elements has exactly seven rows. The Lucas model also predicts the
structure of the nucleus and correctly predicts thousands of
nuclide spins. Boudreaux and Baxter recently have shown that the
Lucas model of the nucleus produces more accurate predictions of
radioactivity and decay rates than prior models \cite{BOU20}.}

\subsection{The Bergman-Lucas Model for Elementary Particles, Atoms and Nuclei}

The abstract of paper \cite{BER97}:{\sl A theory of physical
matter based on fundamental laws of electricity and magnetism is
presented. A new physical model for elementary particles, the atom
and the nucleus implements scientific principles of objective
reality, causality and unity. The model provides the
experimentally observed size, mass, spin, and magnetic moment of
all the stable charged elementary particles. The model is based on
a classical electrodynamics rotating charge ring. From
combinatorial geometry, the complete structure of the Periodic
Table of Elements is predicted, and the nuclear spins and
structure of nuclear shells predicted. Unlike modern mathematical
models based on point-like objects, a physical model has
characteristics of size and structure -- providing a causal
mechanism for forces on objects and the interchange of energy
between objects. From the fundamental laws of electrodynamics and
Galilean invariance, the so-called relativistic fields of a
charged particle moving at high velocity have been derived. The
results are mathematically identical to those predicted by the
Special Theory of Relativity, but the origin of the effect is
entirely physical. The model even accounts for the interaction of
light and matter, and the physical process for absorption and
emission of radiation by an electron is explained from classical
electrodynamics. Using a ring particle absorption mechanism,
classical explanations are given for black body radiation and
photoelectric effect.}

\subsection{ The Hydrogen Atom in Classical Mechanics}

Is it possible to understand some properties of a hydrogen atom
from classical mechanics ? The Hamiltonian for a hydrogen atom is

$$H=\frac{m_p \dot{\vec{r_p}}\;^2}{2} +
\frac{m_e \dot{\vec{r_e}}\;^2}{2} - \frac{e^2}{ \mid \vec{r}_p -
\vec{r}_e \mid }.\eqno(10)$$

All notation is standard. The definition of the center of mass is

$$m_{p}\vec{r}_{p}+m_{e}\vec{r}_{e}=0, \eqno(11) $$

and the relative distance between electron and proton is

$$\vec{r}=\vec{r}_p- \vec{r}_e. \eqno(12)$$

Equations (10)-(12) lead to the results:

$$\vec{r}_{p}=\frac{m_{e}}{m_{p}+m_{e}}\vec{r},\;\vec{r}_{e}=
-\frac{m_{p}}{m_{p}+m_{e}}\vec{r}, \eqno(13)$$

$$H=\frac{\mu \dot{{\vec
r}}\;^2}{2}-\frac{e^{2}}{r},\eqno(14)$$

where
$$\mu=\frac{m_{p}m_{e}}{m_{p}+m_{e}}.\eqno(15)$$

The Hamiltonian (14) coincides with the Hamiltonian for the
fictitious material point with reduced mass $\mu$ moving in the
external field $-e^{2}/r$. If we known the trajectory of this
fictitious particle $\vec{r}=\vec{r}(t)$ then we can reconstruct
the trajectories of electron and proton using equations (13)

$$\vec{r}_{p}(t)=\frac{m_{e}}{m_{p}+m_{e}}\vec{r}(t),\;\;\;
\vec{r}_{e}(t)=-\frac{m_{p}}{m_{p}+m_{e}}\vec{r}(t).\eqno(16)$$

It is evident from (16) that the proton and electron  move in the
opposite directions synchronously. So the motions of proton,
electron and their relative motion occur with  equal frequency

$$\omega_{p}=\omega_{e}=\omega_{\mu},\eqno(17)$$
over the closed trajectories scaling by the ratio

$$ \frac{v_{e}}{v_{p}}=\frac{m_{p}}{m_{e}},\;\frac{v_{e}}{v_{\mu}}=\frac{m_{\mu}}{m_{e}},\;
\frac{v_{\mu}}{v_{p}}=\frac{m_{p}}{m_{\mu}}.\eqno(18)$$

I.A. Schelaev \cite{SCH04} proved that the frequency spectrum of
any motion on ellipse contains only one harmonic.

We can get from (16) that
$$\vec{P}_{p}=\vec{P},\;\vec{P}_{e}=-\vec{P}, \eqno(18a)$$

where -- $\vec{P}_{i}=m_{i}\vec{\dot{r}}_{i}$. All three impulses
are equal to each other in absolute value, which means the
equality of
$$\lambda_{D}(p)=\lambda_{D}(e)=\lambda_{D}(\mu)=h/P.\eqno(19)$$

Conclusion:\\
{\sl $\otimes \;\;\;$ Therefore, the motions of proton and
electron and their relative motion occur with the same FREQUENCY,
IMPULSE (linear momentum) and the de Broglie WAVELENGTH. All
motions are synchronized and self-sustained. Therefore, the whole
system -hydrogen atom  nondecomposable into independent motions of
proton and electron despite the fact that the kinetic energy ratio
of electron to proton  is small:

$$ \frac{E_{k}(e)}{E_{k}(p)}=4.46*10^{-4}.$$
It means that the nuclear and the corresponding atomic processes
must be considered as a unified  entirely determined whole process
as the motions in the Solar system (remember the Moon faces the
Earth without changing its visible side, the same case is in
hydrogen atoms for protons and electrons)}

For example, V.F. Weisskopf \cite{6} came to the conclusion that
the maximum height $H$ of mountains in terms of the Bohr radius
$a$ is equal to

$$\frac{H}{a}=2.6*10^{14},$$
and water wave lengths $\lambda$ on the surface of a lake in terms
of the Bohr radius is equal to

$$\frac{\lambda}{a}\approx 2\pi*10^{7}.$$

$\bullet$ The greatness of mountains, the finger sized drop, the
shiver of a lake, and the smallness of an atom are all related by
simple laws of nature – Victor F. Weisskopf  \cite{6}.

\subsection{The Gareev Model}

Let us introduce the quantity $f=rv$ which is the invariant of
motion, according to the second Kepler law, then

$$ \mu v=\frac{\mu vr}{r}=\frac{\mu f}{r},\eqno(20)$$
and we can rewrite  equation (14) in the following way:

$$ H=\frac{\mu f^{2}}{2r^{2}}-\frac{e^{2}}{r}. \eqno(21)$$

We can obtain the minimal value of (21) by taking its first
derivative over $r$ and setting it equal to zero. The minimal
value occurs at

$$ r_{0}=\frac{\mu f^{2}}{e^{2}}, \eqno(22)$$
and the result is

$$ H_{min}=E_{min}=-\frac{e^{4}}{2 \mu f^{2}}.\eqno(23)$$

The values of invariant of motion  $\mu f$ (in MeV*s) can be
calculated from (23) if we require the equality of  $E_{min}$ to
the energy of the ground  state of a hydrogen atom

$$ \mu f= \mu
vr=6.582118*10^{-22}=\hbar,\;\eqno(24)$$

Conclusion:\\
{\sl $\otimes \;\;\;$ The Bohr quantization conditions were
introduced as a hypothesis. We obtain these conditions from a
classical Hamiltonian requiring its minimality. It is necessary to
strongly stress that no assumption was formulated about
trajectories of proton and electron. We reproduced exactly the
Bohr result and modern quantum theory. The Plank constant $\hbar$
is the Erenfest adiabatic invariant for a hydrogen atom: $\mu vr =
\hbar$.}

Let us briefly review our steps:

$\bullet$ We used a well established interaction between proton
and electron.

$\bullet$ We used a fundamental fact that the total energy=kinetic
energy+potential energy.

$\bullet$ We used the second Kepler law.

$\bullet$ We used usual calculus to determine the minimum values
of $H$.

$\bullet$ We required the equality of  $E_{min}$ to the energy of
the ground state of hydrogen atom.

 Classical Hamiltonian + classical interaction between proton and
electron + classical second Kepler law + standard variational
calculus -- these well established steps in macrophysics reproduce
exactly  results of the Bohr model and modern quantum theory
(Schrodinger equation) -- results of microphysics. We have not
done anything spectacular or appealed to any revolutionary and
breakthrough physics.

\subsection{The Gryzinski model}

In this subsection we shortly highlighted  a very important
results (which were entirely ignored and forgotten despite that
these papers were published in famous peer-reviewed journals)
obtained by M. Gryzinski \cite{GRY04} on the basis of the Newton
equation with well established Coulomb interactions. M. Grysinski
pointed out that there was a lot of arguments that classical
dynamics at the atomic level work, and that the concept of a
localized electron was abandoned too early. It is a very
interesting to bring some of his quotations
(http://www.iea.cyf.gov.pl/gryzinski/misiek.html):

$\bullet$ {\sl Since the time Bohr  formulated his famous
correspondence principle questioning applicability of classical
dynamics to description of atomic system, and Heisenberg  spread
an electron  in space by his famous inequality, dynamical
considerations initiated by works of Thomson \cite{THO06} and
Rutherford \cite{RUT11} have disappeared from atomic physics
almost completely. It was a result of a highly restrictive form of
both the principles.

The author ignoring these principles turned back in 1957 to the
old idea of a localized electron and showed that the classical
collision theory developed on the basis of a classical two body
problem worked \cite{GRY57}.

On the basis of the Newton equation of motion and Coulomb low
there were accurately described:

1. Collisional ionization and excitation of atoms and molecules
\cite{GRY65},

2. Ramsauer effect and Vander Waals forces \cite{GRY70,GRY75},

3. Atomic diamagnetism \cite{GRY87a} and atomic energy level
shifts \cite{GRY76},

4. Electronic structure of $He^{+}_{2}$ \cite{GRY87} as well as
dynamical nature of a covalent bond \cite{GRY94}.}

 M. Gryzinski \cite{GRY04} proved
that atoms have the quasi-crystal structure with definite angles:
$90^{\circ}$, $109^{\circ}$ and $120^{\circ}$ which are the
well-known angles in crystallography.

\subsection{The Gudim-Andreeeva Model}

Authors of \cite{GUD01,GUD03,GUD06} propose a classical procedure
to calculate the potential energy of electrons in the ground state
of atoms using the interaction between an electron and a proton in
the form

$$V=-\frac{e^{2}}{r}+\frac{\hbar^{2}}{2mr^{2}},\eqno(g1).$$
They were able to calculate the ground state energies for six
lightest atoms in  the reasonable agreement with experimental
data.

It is well-known that the electron  trajectories in the Kepler
problem with the Coulomb potential for the finite motion are
represent the closed orbits for any energy. Note that the closed
orbits for the binomial potential (g1) exist only for discrete
values of energy \cite{GUD01}. The Bohr model use the Coulomb
force plus centrifugal one which means that an electron rotate
around the nucleus. This assumption leads to the difficulties in
the interpretations of the experimental data for the hydrogen
atom.
\subsection{The Huang Model }

We should like to highlight of results obtained by X. Q. Huang
\cite{HUA06} using the classical electromagnetic field theory.\\

$\bullet$ X.Q. Huang wrote \cite{HUA06}:\\
{\sl We study the energy conversion laws of the macroscopic
harmonic $LC$ oscillator, the electromagnetic wave (photon) and
the hydrogen atom. As our analysis indicates that the energies of
these apparently different systems obey exactly the same energy
conversion law. Based on our results and the wave-particle duality
of electron, we find that the atom in fact is a natural
microscopic $LC$ oscillator. In the framework of classical
electromagnetic field theory we analytically obtain, for the
hydrogen atom, the quantized electron atom orbit radius
$r_{n}=a_{0}n^{2}$, and quantized energy $E_{n}=-R_{H}hc/n^{2},
(n=1,2,3,..)$, where $a_{0}$ is the Bohr radius and $R_{H}$ is the
Rydberg constant. Without any adaptation of the quantum theory, we
present a reasonable explanation of the polarization of photon,
the Zeeman effect, Selection rules and Pauli exclusion principle.
Our results show that the concept of electron spin is not the
physical reality and should be replaced by the intrinsic
characteristic of the helical moving electron (Left-hand and
Right-hand). In addition, a possible physical mechanism of
superconductivity and a deeper physical understanding of the
electron mass are also provided.}

 X.Q. Huang considered in first time the hydrogen atom as a
 natural microscopic $LC$ oscillator and he obtained the  results in excellent
 agreement with the Bohr model and quantum mechanical theory.

\subsection{The Mills Model }

The conventional point of view  is that the validity of the
Maxwell equations is restricted only to the macroscale and that
they do not apply to the atomic scale. R.L. Mills \cite{MIL05}
developed the model of atoms on the Maxwell equation which he
called "The Grand Unified Theory of Classical Quantum Mechanics"
(CQM). Under special conditions, an extended distribution of
charge may accelerate without radiation energy. The mathematical
formulation for zero radiation based on Maxwell's equations
follows from a derivation by Haus\cite{HAU86}. This leads to a
physical model of subatomic particles, atoms, and molecules.

Equations are closed-form solutions containing fundamental
constants only and agree with experimental observations. The
calculated energies from exact solutions of one through
twenty-electron atoms are available from the internet \cite{MIL}.

$\bullet$ R.L. Mills came to the conclusion: {\sl for 400 atoms
and ions the agreement between the predicted and experimental
results is remarkable. Other problems exactly solved as further
tests of CQM are the anomalous magnetic moment of the electron,
the Lamb Shift, the fine structure and superfine structure of the
hydrogen atom, the superfine structure intervals of positronium
and muonium. The agreement between observations and predictions
based on closed-form equations with fundamental constants only
matches  the limit permitted by the error in the measured
fundamental constants.

The solution of the nature of the electron and photon for the
first time also allow for exact solutions of excited states. For
ever 100 excited states  of the helium atom, the r-squared value
is 0.999994, and the typical average relative difference is about
5 significant figures which is within the error of the
experimental data.. Using only the Coulomb energy at the
calculated radii, the agreement is remarkable. These results
demonstrate the predictive power of CQM that further provides the
nature of and conditions to form lower-energy states of hydrogen
which are also based on electron-photon interactions.}

$\bullet$ Conclusion: Splendidly, all atomic models  based on
either the Newton equation and the Kepler laws or  the Maxwell
equations, or  the Schrodinger and Dirac equations achieved
agreement with experimental data.

We can only suspect that these equations are grounded on the same
fundamental principle(s) which is (are) not known or these
equations can be transformed into each other. R.D. Feynman
\cite{DYS90} proved the Maxwell equations assuming only Newton's
law of motion and the commutation relation between the position
and velocity for a single nonrelativistic particle. The Dirac
equation can be rewritten in the Maxwell equations form
\cite{BUL94}.

Bohr and Schrodinger assumed that the laws of physics that are
valid in the macrosystem do not hold in the microworld of the
atom. We think that the laws in macro- and microworld are the
same.

\section{Nuclei and Atoms as Open Systems}

1) LENR may be understood in terms of the known fundamental laws
without any violation of the basic physics. The fundamental laws
of physics should be  the same in micro- and macrosystems.

2)Weak and electromagnetic interactions may show a strong
influence of the surrounding conditions on the nuclear
processes.\\

3)The conservation laws are valid for  closed systems. Therefore,
the failure of parity in weak interactions means that the
corresponding systems are  open systems. Periodic variations (24
hours, 27, and 365 days in  beta-decay rates indicate that the
failure of parity in weak interactions has a cosmophysical origin.
Modern quantum theory is the theory for closed systems. Therefore,
it should be reformulated for open systems. The closed systems are
idealization of nature,  they do not exist in reality. \\

4)The universal cooperative resonance synchronization principle
is a key issue to make a bridge between various scales of
interactions and it is responsible for self-organization of
hierarchical systems independent of substance, fields,  and
interactions. We give some arguments in favor of the mechanism –
ORDER BASED on ORDER, declared  by Schrodinger in \cite{4}, a
fundamental problem of contemporary science.\\

5)The universal resonance synchronization principle became a
fruitful interdisciplinary science of general laws of
self-organized processes in different branches of physics because
it is the consequence of the energy conservation law and resonance
character of any interaction between wave systems. We have proved
the homology of  atom, molecule and crystal structures including
living cells. Distances of these systems are commensurable  with
the de Broglie wave length  of an electron in the ground state of
a hydrogen atom,  it plays the role of the standard distance, for
comparison. \\

6)First of all, the structure of a hydrogen atom should be
established. Proton and electron in a hydrogen atom move with the
same frequency that creates attractive forces between them, their
motions are synchronized. A hydrogen atom represents the radiating
and accepting  antennas (dipole) interchanging  energies with the
surrounding  substance. The sum of radiated and absorbed energy
flows by electron and  proton in a stable orbit is equal to zero
\cite{5} – the secret of success of  the Bohr model (nonradiation
of  the electron on a stable orbit). “The greatness of mountains,
the finger sized drop, the shiver of a lake, and the smallness of
an atom are all related by simple laws of nature” – Victor F. Weisskopf  \cite{6}.\\

7)These flows created  standing waves due to the cooperative
resonance synchronization principle. A constant energy exchange
with substances (with universes) create stable auto-oscillation
systems in which the frequencies of  external fields and all
subsystems are commensurable. The relict radiation (the relict
isotropic standing waves at T=2.725 K – the Cosmic Microwave
Background Radiation (CMBR))  and   many isotropic standing waves
in cosmic medium \cite{7} should be results of self-organization
of the stable atoms, according to the universal cooperative
resonance synchronization principle that is a consequence of  the
fundamental energy conservation law. One of the fundamental
predictions of the Hot Big Bang theory for the creation of the
Universe is CMBR.\\

8)The cosmic isotropic standing waves (many of them are not
discovered yet) should play the role of a conductor responsible
for stability of elementary particles, nuclei, atoms,…, galaxies
ranging in size more than 55 orders of magnitude.\\

9)The phase velocity of standing microwaves can be extremely high;
therefore, all objects of the Universe should  get information
from each other almost immediately using phase velocity.

The aim of this paper is to discuss the possibility of inducing
and controlling nuclear reactions at low temperatures and
pressures by using different low energy external fields and
various physical and chemical processes. The main question is the
following: is it possible to enhance LENR rates by using  low and
extremely low energy external fields? The review of possible
stimulation mechanisms is presented in \cite{2,5}. We will discuss
new  possibilities to enhance LENR rates in condensed matter.

\section{LENR in Condensed Matter}

Modern understanding of the decay of the neutron is

$$n \rightarrow p+e^{-}+\overline{\nu}_{e}.\eqno(25)$$
The energetics  of the decay can be analyzed using the concept of
binding energy and the masses of  particles by their rest mass
energies. The energy balance from neutron decay can be calculated
from the particle masses. The rest mass difference ( $0.7823
MeV/c^{2}$) between neutron and (proton+electron) is converted to
the kinetic energy of proton, electron, and neutrino. The neutron
is about $0.2\%$ more massive than a proton, a mass difference is
1.29 $MeV$. A free neutron will decay with a half-life of about
10.3 minutes. Neutron in a nucleus will decay if a more stable
nucleus results; otherwise neutron in a nucleus will be stable. A
half-life of a neutron in nuclei changes dramatically and depends
on the isotopes.

The capture of electrons by protons

$$p+e^{-}\rightarrow n+\nu_{e},\eqno(26)$$
but for free protons and electrons this reaction has never been
observed which is the case in nuclear+ atomic physics. The capture
of electrons by protons in a nucleus will occurs if a more stable
nucleus results. \\

\subsection{Cooperative Processes}

The processes (25) and (26) in LENR  are going with individual
nucleons and electrons. In these cases the rest mass difference is
equal to $0.7823 MeV/c^{2}$. In the case of neutron decay the
corresponding energy ($Q=0.7823$ MeV) converted to kinetic
energies of proton, electron, and antineutrino. In the case of the
capture of electrons by protons the quantity  $Q=0.7823$ MeV is a
threshold electron kinetic energy under which the process (26) is
forbidden for free proton and electron.

We have formulated the following postulate:\\
$\otimes$ {\sl The processes (25) and (26) in LENR  are going  in
the whole system: cooperative processes including all nucleons in
nuclei and electrons in atoms, in condensed matter. In these cases
a threshold energy $Q$ can be drastically decreased by internal
energy of the whole system or even more -- the electron capture by
proton can be accompanied by emission of internal binding energy -
main source of excess heat phenomenon in LENR. }

The processes (25) and (26) are weak processes. A weak interaction
which is responsible for electron capture and other forms of beta
decay  is of a very short range. So the rate of electron capture
and emission (internal conversion) is proportional to the density
of electrons in nuclei. It means that we can manage the
electron-capture (emission) rate by the change of the total
electron density in the nuclei using different low energy external
fields. These fields  can play a role of triggers for extracting
internal energy of the whole system or subsystems, changing
quantum numbers of the initial states in such a way that forbidden
transitions become allowed ones. The distances between proton and
electron in atoms are  of the order $10^{-6}--10^{-5}$ cm and any
external field decreasing these distances even for a small value
can increase the process (26) in nuclei in an exponential way.
Therefore, the influence of an external electron flux (discharge
in condensed matter: breakdown, spark and ark) on the velocity
processes (25) and (26) can be of great importance.

The role of external electrons is the same as the catalytic role
of neutrons in the case of the chain fission reactions in nuclei
-- neutrons bring to nuclei binding energies (about 8 MeV) which
enhance the fission rates by about 30 orders.

\section{Predicted Effects and Experimentum  Crucis}

Postulated enhancement mechanism of LENR by external fields can be
verified by the Experimentum Crucis. We \cite{5} predicted that
natural geo-transmutation in the atmosphere and earth occurs in
the regions of a strong change in geo-, bio-, acoustic-,...  and
electromagnetic fields.

Various electrodynamic processes at thunderstorms are responsible
for different phenomena: electromagnetic pulses, $\gamma$-rays,
electron fluxes, neutron fluxes, and radioactive nuclei fluxes.

\subsection{Neutron Production by Thunderstorms}

 The authors of  \cite{BRA4} concluded that a  neutron burst is
associated with lighting. The total number of neutrons produced by
one typical lightning discharge was estimated as $2.5*10^{10}$.

\subsection{Production of Radiocarbon and Failing of Radiocarbon
Dating}

The radiocarbon dating is based on the decay rate of radioactive
isotope $^{14}C$ which is believed to be constant irrespective of
the physical and chemical conditions. The half-life of radiocarbon
$^{14}C$ is 5730 years. A method for historical chronometry was
developed assuming that the decay ratio of $^{14}C$ and its
formation are constant in time. It was postulated that $^{14}C$ is
formed only by the cosmic ray neutrons

$$^{14}N(n,p)^{14}C. \;\eqno(27)$$
Radiocarbon dating is widely used in archeology, geology,
antiquities,... There are over 130 radiocarbon dating
laboratories. The radiocarbon method of dating was developed by
Willard F. Libby who was awarded  the Nobel prize in Chemistry for
1960.

The radiocarbon method does not take into account the following
facts which have been established recently:

$\otimes$ The neutron production by thunderstorms \cite{BRA4}

$\otimes$ The Production of radiocarbon by lighting bolts
\cite{LIB73}.

 Let us consider the reaction

$$^{14}_{7}N+e^{-}\rightarrow ^{14}_{6}C+\nu_{e},\eqno(27a)$$
$T_{k}(e)$=156.41 keV is the threshold energy which should by
compared with 782.3 keV for process (26). Production of
radiocarbon by lighting bolts was established in \cite{LIB73}.

\subsection{Production Radiophosphorus by Thunderstorms}

The life-times of $^{32}_{15}P$ and $^{33}{15}P$ are equal to
14.36 and 25.34 days, respectively. They were found in rain-water
after thunderstorms \cite{SEL70}. Production of the
radiophosphorus by thunderstorms can be understood in the
following way:

$$^{32}_{16}S+e^{-}\rightarrow ^{32}_{15}P+\nu_{e},\eqno(28)$$
$$^{33}_{16}S+e^{-}\rightarrow ^{33}_{15}P+\nu_{e},\eqno(29)$$
thresholds of these processes are equal to 1.710 and 0.240 MeV,
respectively.  The precipitation of MeV electrons from the inner
radiation belt \cite{INAN} and enhancement of the processes by
lightning are possible.

\subsection{LENR Stimulated by Condensed Matter Discharge}

Let us consider the condensed matter discharge (breakdown, spark
and arc) using the
different electrode.  There are the following processes:\\

1. The electrode is $Ni$. Orbital or external electron capture
$$^{58}_{28}Ni(68.27\%)+e^{-}\rightarrow ^{58}_{27}Co(70.78\;
days)+\nu_{e}, \eqno(30)$$

The threshold $Q_{1}=0.37766\;keV$ of this reaction on $Ni$ should
be compared with the threshold $Q_{2}=0.7823$ energy for electron
capture by free protons: $Q_{2}/Q_{1}\approx 2$. The velocity of
orbital electron capture can be enhanced by the discharge.

2.Orbital or external electron capture
$$^{58}_{27}Co(70.78\; days)+e^{-}\rightarrow ^{58}_{26}Fe(0.28\%)+\nu_{e},\eqno(31)$$
with emission of energy $Q_{2}=2.30408$ MeV.

3. Double orbital or external electron capture
$$^{58}_{28}Ni(68.27\%)+2e^{-} \rightarrow ^{58}_{26}Fe(0.28\%)+2\nu_{e},\eqno(32)$$
with emission of energy $Q_{3}=1.92642$ mostly by neutrinos.

The proposed cooperative mechanism of LENR in this case can be
proved in an extremely simple way: presence of radioactive
$^{58}_{27}Co$ and enriched isotope of $^{58}_{26}Fe$.

$\otimes$ {\sl This mechanism can give  possibilities to get a way
of controlling  the necessary isotopes and excess heat.}

\subsection{Neutrinoless Double Beta Decay}

As we known \cite{2}, the physical roles of electron and neutrino
for LENR in condensed matter has not been investigated in detail
up to now despite the fact that  weak processes in nuclei are well
understood. The double beta decay is the rarest spontaneous
nuclear transition,in which the nuclear charge changes by two
units while the mass number remains the same. Such a case can
occur for the isobaric triplet $A(Z,N)$, $A(Z\pm 1,N\mp 1)$,
$A(Z\pm 2,N\mp 2)$, in which the middle isobar has a greater rest
mass than the extreme ones, and the extremes are the nuclei with
the even $Z$ and $N$. The usual beta-decay transferring a given
nucleus into another via an intermediate nucleus is energetically
forbidden.

The double beta decay in nuclei  can proceed in different modes \cite{KLA95}: \\

$\otimes$The two neutrinos decay mode $2\nu\beta\beta$ is

$$A(Z,N)\rightarrow A(Z+2,N-2)+2e^{-}+2\overline{\nu}_{e},\eqno(33)$$
which is allowed by the Standard Model of particle physics. The
total kinetic energy of two emitted electrons present  continuous
spectra up to $E_{max}$.

$\otimes$The neutrinoless mode $0\nu\beta\beta$

$$A(Z,N)\rightarrow A(Z+2,N-2)+2e^{-},\eqno(34)$$
which requires violation of a lepton number. The total kinetic
energy of two emitted electrons is equal to $E_{max}$.

Two neutrinos in the mode $2\nu\beta\beta$ carry out almost all
emitted energies.  A fundamental question is: Does the
neutrinoless double beta decay exist or not (for the review of the
history see  \cite{KLA95,KLA05})?. The emerged energies in the
neutrinoless $0\nu\beta\beta$ mode are easily  detected for
practical use but these are the rarest spontaneous nuclear
transitions ($T\approx 10^{18}-10^{30}$ years). Is it possible to
enhance the decay rate?

Above and in \cite{1,2,3} we have discussed the cooperative and
resonance synchronization enhancement mechanisms of LENR. Some of
the low energy external fields can be used as triggers for
starting and enhancing  exothermic LENR. It is natural to expect
that in the case of beta-decay (capture)  the external electron
flux with high density, or the laser of high intensity, or any
suitable external fields should play this  role. Any external
field shortening distances between protons in nuclei and electrons
in atoms should enhance beta-decay (capture) or double-beta decay
(capture).

There is a great number of experiments in Japan, Italy, Russia,
US, India, China, Israel, and Canada in which cold transmutations
and excess energy were measured (see http://www.lenr-canr.org).
Indeed the existence of LENR is now well established but the
proposed about 150 theoretical models for interpretation of
experimental data are not accepted (A. Takahashi, ICCF12).

It is  very popular to use $Ni$, $Pd$, $Pt$ and $W$ as electrodes
in the condensed matter discharge (breakdown, spark, arc, and
explosion) experiments. Let us consider the case of $Pd$
electrodes. The difference of the rest masses are equal
$$m(^{110}_{46}Pd)-m(^{110}_{48}Cd)=1.9989\; MeV/c^{2};$$
therefore, the external field can open the channel
$^{110}_{46}Pd\rightarrow ^{110}_{48}Cd$ with $Q=1.9989$ MeV. In
the cases $Ni$, $Pt$, and $W$ we have

$$m(^{58}_{28}Ni)-m(^{58}_{26}Fe)=1.92642\;MeV/c^{2},$$
$$m(^{186}_{74}W)-m(^{186}_{76}Os)=0.47302\;MeV/c^{2},$$
$$m(^{198}_{78}Pt)-m(^{198}_{80}Hg)=1.05285\;MeV/c^{2}.$$

The proposed cooperative mechanism of LENR in these cases can be
proved in an extremely simple way: presence of  enriched isotopes
of $^{58}_{26}Fe$, $^{110}_{48}Cd$, $^{186}_{76}Os$, and
$^{198}_{80}Hg$ for the indicated above electrodes.

The experimental data \cite{SAV96,KRI03,KUZ03,KRY02,BAL03} seem to
confirm such expectations.

Therefore,  expensive and time consuming  double beta decay
experiments can be performed in extremely  cheap and short-time
experiments by using suitable external fields. This new direction
of research can give answers for fundamental problems of modern
physics (the lepton number conservation, type of neutrino,
neutrino mass spectrum,... ), it can open production of new
elements (utilization of radioactive waste) and excess heat
without an ecological problem.

 A careful analysis of the double
beta decay shows that the $2e^{-}$ cluster can be responsible for
the double beta decay. The difference between the rest mass
$^{130}_{56}Ba$ and $^{130}_{52}Te$, which is equal to 92.55 keV,
indicates the possibilities to capture the $4e^{-}$ cluster by
$^{130}_{56}Ba$. It is a full analogy with the Iwamura reactions
\cite{IWA11}.

\section{Quantization of Nuclear and Atomic Rest Masses}

Almost all quantomechanical models describe excited states of
nuclei, atoms, molecules, condensed matter,... neglecting
structure of the ground state of the investigated systems.
Therefore, we have  very restricted information about the
properties of nuclei, atoms,... in their ground states. Note that
the mutual influence of the nucleon and electron spin (the
superfine splitting), the Mossbauer effect,... are well-known. The
processes going in the surrounding  matter of nuclei change the
nuclear moments and interactions of nucleons in nuclei.

We proved  that the motions of proton and electron in the hydrogen
atom in the ground state occur with the same frequency; therefore,
their motions are synchronized. The cooperation in motion of
nucleons in nuclei and electrons in atoms in their ground states
is still an open problem for  nuclei and atoms having many
nucleons and electrons, respectively.

$\bullet$ {\sl We formulate a very simple and audacious working
hypothesis: the nuclear and the corresponding atomic processes
must be considered as a unified entirely determined whole process.
The nucleons in nuclei and the electrons in atoms form  open
nondecomposable whole systems in which all frequencies and phases
of nucleons and electrons are coordinated according to the
universal cooperative resonance synchronization principle.}

This hypothesis can be proved at least partly by investigation of
the difference between nuclear and atomic rest masses. We
performed this analysis for the first time, experimental data from
\cite{MOL97}.

\newpage
\begin{center}
Table 1.    \\
The differences between nuclear and atomic rest masses
$\Delta M(Z,A)=M_{atom}(Z,A)-M_{nuclei}(Z,A)$,\;
$\Delta=\Delta M(Z,A)-\Delta M(Z,A-1)$, in \; $MeV/c^{2}$\\

\vspace*{0.2cm}
\begin{tabular}{||c|c|c|c|c|c|c|c||}
\hline Nuclei & $\Delta M$ &$\Delta$& Nuclei&$\Delta M$&$\Delta$&$\delta$&$\delta\delta$  \\
\hline
$^{1}_{1}H$&0.51099&       &            &       &       &        &\\
$^{2}_{1}H$&0.51105&0.00006&$^{3}_{2}He$&1.02187&       &0.51082 &\\
$^{3}_{1}H$&0.51117&0.00012&$^{4}_{2}He$&1.02210&0.00023&0.51093&0.00011\\
$^{4}_{1}H$&0.51093&0.00024&$^{5}_{2}He$&1.02186&0.00024&0.51093&0.00000\\
$^{4}_{3}Li$&1.53327&      &$^{6}_{2}He$&1.02186&0.00000&0.51141&0.00048 \\
$^{5}_{3}Li$&1.53351&0.00024&$^{7}_{2}He$&1.02186&0.00000&0.51165&0.00024\\
$^{6}_{3}Li$&1.53255&0.00096&$^{8}_{2}He$&1.02186&0.00000&0.51069&0.00096\\
$^{7}_{3}Li$&1.53350&0.00095&$^{9}_{2}He$&1.02139&0.00047&0.51211&0.00142\\
$^{8}_{3}Li$&1.53351&0.00001&$^{10}_{2}He$&1.02138&0.00001&0.51213&0.00002\\
$^{9}_{3}Li$&1.53256&0.00095&             &       &       &       &\\
\hline
\end{tabular}
\end{center}

\begin{center}
Table (continued)    \\

\vspace*{0.2cm}
\begin{tabular}{||c|c|c|c|c|c|c|c||}
\hline Nuclei & $\Delta M$ &$\Delta$& Nuclei&$\Delta M$&$\Delta$&$\delta$&$\delta\delta$  \\
\hline
$^{6}_{4}Be$&2.04420&       &$^{7}_{5}B$&2.55537&       &0.51117 &\\
$^{8}_{4}Be$&2.04468&0.00048&$^{8}_{5}B$&2.55537&0.00000&0.51069&0.00048\\
$^{9}_{4}Be$&2.04468&0.00000&$^{9}_{5}B$&2.55584&0.00047&0.51116&0.00047\\
$^{10}_{4}Be$&2.04468&0.00000&$^{10}_{5}B$&2.55489&0.00095&0.51021&0.00095 \\
$^{11}_{4}Be$&2.04467&0.00001&$^{11}_{5}B$&2.55584&0.00047&0.51117&0.00096\\
$^{12}_{4}Be$&2.04373&0.00094&$^{12}_{5}B$&2.55584&0.00000&0.51211&0.00094\\
$^{13}_{4}Be$&2.04372&0.00001&$^{13}_{5}B$&2.55585&0.00001&0.51213&0.00002\\
$^{14}_{4}Be$&2.04467&0.00095&$^{14}_{5}B$&2.55489&0.00096&0.51022&0.00191\\
             &       &       &$^{15}_{7}B$&2.55585&0.00096&       &  \\

\hline
\end{tabular}
\end{center}

\newpage
\begin{center}
Table 1 (continued)    \\

\vspace*{0.2cm}
\begin{tabular}{||c|c|c|c|c|c|c|c||}
\hline Nuclei & $\Delta M$ &$\Delta$& Nuclei&$\Delta M$&$\Delta$&$\delta$&$\delta\delta$  \\
\hline
$^{8}_{6}C$&3.06702&       &$^{11}_{7}N$&3.57914&       &       &        \\
$^{9}_{6}C$&3.06702&0.00000&$^{12}_{7}N$&3.57819&0.00095&0.51117&\\
$^{10}_{6}C$&3.06701&0.00001&$^{13}_{7}N$&3.57819&0.00000&0.51118&0.00001\\
$^{11}_{6}C$&3.06702&0.00001&$^{14}_{7}N$&3.57818&0.00001&0.51116&0.00002 \\
$^{12}_{6}C$&3.06702&0.00000&$^{15}_{7}N$&3.57818&0.00000&0.51116&0.00000\\
$^{13}_{6}C$&3.06702&0.00000&$^{16}_{7}N$&3.58819&0.00001&0.51117&0.00001\\
$^{14}_{6}C$&3.06702&0.00000&$^{17}_{7}N$&3.57818&0.00001&0.51116&0.00001\\
$^{15}_{6}C$&3.06701&0.00001&$^{18}_{7}N$&3.57818&0.00000&0.51117&0.00001\\
$^{16}_{6}C$&3.06702&0.00001&$^{19}_{7}N$&3.57818&0.00000&0.51116&0.00001  \\
$^{17}_{6}C$&3.06702&0.00000&$^{20}_{7}N$&3.57628&0.00190&0.50926&0.00190 \\
$^{18}_{6}C$&3.06893&0.00191&$^{21}_{7}N$&3.57818&0.00190&0.50925&0.00001\\
            &       &       &$^{22}_{7}N$&3.57819&0.00001&       &\\
\hline
\end{tabular}
\end{center}

\begin{center}
Table 1 (continued)    \\

\vspace*{0.2cm}
\begin{tabular}{||c|c|c|c|c|c|c|c||}
\hline Nuclei & $\Delta M$ &$\Delta$& Nuclei&$\Delta M$&$\Delta$&$\delta$&$\delta\delta$  \\
\hline
$^{12}_{8}O$&4.09031&       &$^{15}_{9}F$&4.60148&       &0.51117& \\
$^{13}_{8}O$&4.09031&0.00000&$^{16}_{9}F$&4.60243&0.00095&0.51212&0.00095\\
$^{14}_{8}O$&4.08936&0.00095&$^{17}_{9}F$&4.60147&0.00096&0.51211&0.00001\\
$^{15}_{8}O$&4.09031&0.00095&$^{18}_{9}F$&4.60244&0.00097&0.51213&0.00002 \\
$^{16}_{8}O$&4.09031&0.00000&$^{19}_{9}F$&4.60243&0.00001&0.51212&0.00001\\
$^{17}_{8}O$&4.09031&0.00000&$^{20}_{9}F$&4.60243&0.00000&0.51212&0.00000\\
$^{18}_{8}O$&4.08936&0.00095&$^{21}_{9}F$&4.60434&0.00191&0.51498&0.00286\\
$^{19}_{8}O$&4.08936&0.00000&$^{22}_{9}F$&4.60052&0.00382&0.51116&0.00382\\
$^{20}_{8}O$&4.09031&0.00095&$^{23}_{9}F$&4.60053&0.00001&0.51022&0.00094  \\
$^{21}_{8}O$&4.09126&0.00095&$^{24}_{9}F$&4.60053&0.00000&0.50927&0.00095 \\
$^{22}_{8}O$&4.08936&0.00190&$^{25}_{9}F$&4.60052&0.00001&0.51116&0.00189\\
\hline
\end{tabular}
\end{center}

\begin{center}
Table 1 (continued)    \\

\vspace*{0.2cm}
\begin{tabular}{||c|c|c|c|c|c|c|c||}
\hline Nuclei & $\Delta M$ &$\Delta$& Nuclei&$\Delta M$&$\Delta$&$\delta$&$\delta\delta$  \\
\hline
$^{16}_{10}Ne$&5.11360&       &$^{19}_{11}Na$&5.62477&       &0.51117 \\
$^{17}_{10}Ne$&5.11360&0.00000&$^{20}_{11}Na$&5.62668&0.00191&0.51308&0.00191\\
$^{18}_{10}Ne$&5.11361&0.00001&$^{21}_{11}Na$&5.62477&0.00191&0.51116&0.00192\\
$^{19}_{10}Ne$&5.11361&0.00000&$^{22}_{11}Na$&5.62477&0.00000&0.51116&0.00000 \\
$^{20}_{10}Ne$&5.11360&0.00001&$^{23}_{11}Na$&5.62477&0.00000&0.51117&0.00001\\
$^{21}_{10}Ne$&5.11550&0.00190&$^{24}_{11}Na$&5.62478&0.00001&0.50928&0.00189\\
$^{22}_{10}Ne$&5.11360&0.00190&$^{25}_{11}Na$&5.62477&0.00001&0.51117&0.00189\\
$^{23}_{10}Ne$&5.11360&0.00000&$^{26}_{11}Na$&5.62667&0.00190&0.51307&0.00190\\
$^{24}_{10}Ne$&5.11169&0.00191&$^{27}_{11}Na$&5.62478&0.00189&0.51309&0.00002 \\
$^{25}_{10}Ne$&5.11169&0.00000&$^{28}_{11}Na$&5.62667&0.00189&0.51498&0.00189\\
$^{26}_{10}Ne$&5.11170&0.00001&$^{29}_{11}Na$&5.62667&0.00001&0.51499&0.00190\\
\hline
\end{tabular}
\end{center}
\newpage
\begin{center}
Table 1 (continued)    \\

\vspace*{0.2cm}
\begin{tabular}{||c|c|c|c|c|c||}
\hline Nuclei & $\Delta M$ &$\Delta$& Nuclei&$\Delta M$&$\Delta$  \\
\hline
$^{20}_{12}Mg$&6.13594&       &$^{24}_{14}Si$&7.16018&        \\
$^{21}_{12}Mg$&6.13594&0.00000&$^{25}_{14}Si$&7.16018&0.00000\\
$^{22}_{12}Mg$&6.13785&0.00191&$^{26}_{14}Si$&7.16209&0.00191\\
$^{23}_{12}Mg$&6.13785&0.00000&$^{27}_{14}Si$&7.16210&0.00001 \\
$^{24}_{12}Mg$&6.13785&0.00000&$^{28}_{14}Si$&7.16209&0.00001\\
$^{25}_{12}Mg$&6.13785&0.00000&$^{29}_{14}Si$&7.16209&0.00000\\
$^{26}_{12}Mg$&6.13785&0.00000&$^{30}_{14}Si$&7.16210&0.00001\\
$^{27}_{12}Mg$&6.13594&0.00191&$^{31}_{14}Si$&7.16209&0.00001\\
$^{28}_{12}Mg$&6.13784&0.00190&$^{32}_{14}Si$&7.16210&0.00001  \\
$^{29}_{12}Mg$&6.13594&0.00190&$^{33}_{14}Si$&7.16019&0.00191 \\
$^{26}_{12}Mg$&6.13594&0.00000&              &        &\\
\hline
\end{tabular}
\end{center}

\begin{center}
Table 1 (continued)    \\
\vspace*{0.2cm}
\begin{tabular}{||c|c|c|c|c|c|c|c||}
\hline Nuclei & $\Delta M$ &$\Delta$& Nuclei&$\Delta M$&$\Delta$&$\delta$&$\delta\delta$ \\
\hline
$^{28}_{15}P$&7.67517&       &$^{28}_{16}S$&8.18824&       &0.51307&\\
$^{29}_{15}P$&7.67517&0.00000&$^{29}_{16}S$&8.18634&0.00190&0.51117&0.00190\\
$^{30}_{15}P$&7.67517&0.00000&$^{30}_{16}S$&8.18634&0.00000&0.51117&0.00000\\
$^{31}_{15}P$&7.67326&0.00191&$^{31}_{16}S$&8.18634&0.00000&0.51308&0.00191 \\
$^{32}_{15}P$&7.67327&0.00001&$^{32}_{16}S$&8.18634&0.00000&0.51307&0.00001\\
$^{33}_{15}P$&7.67327&0.00000&$^{33}_{16}S$&8.18634&0.00000&0.51307&0.00000\\
$^{34}_{15}P$&7.67326&0.00001&$^{34}_{16}S$&8.18634&0.00000&0.51307&0.00000\\
$^{35}_{15}P$&7.67135&0.00191&$^{35}_{16}S$&8.18634&0.00000&0.51499&0.00192\\
$^{36}_{15}P$&7.67517&0.00382&$^{36}_{16}S$&8.18634&0.00000&0.51117&0.00382\\
$^{37}_{15}P$&7.67135&0.00382&$^{37}_{16}S$&8.19016&0.00382&0.51881&0.00764\\
$^{38}_{15}P$&7.67135&0.00000&$^{38}_{16}S$&8.18252&0.00764&0.51117&0.00764\\
$^{39}_{15}P$&7.67517&0.00382&$^{39}_{16}S$&8.19015&0.00763&0.51498&0.00381\\
$^{40}_{15}P$&7.67136&0.00381&$^{40}_{16}S$&8.18634&0.00381&0.51498&0.00000\\
$^{41}_{15}P$&7.67899&0.00763&$^{41}_{16}S$&8.18634&0.00000&0.50735&0.00763\\
\hline
\end{tabular}
\end{center}

\begin{center}
Table 1 (continued)    \\

\vspace*{0.2cm}
\begin{tabular}{||c|c|c|c|c|c|c|c||}
\hline Nuclei & $\Delta M$ &$\Delta$& Nuclei&$\Delta M$&$\Delta$&$\delta$&$\delta\delta$ \\
\hline
$^{85}_{40}Zr$&20.54596&       &$^{85}_{41}Nb$&21.05713&       &0.51117&\\
$^{86}_{40}Zr$&20.53070&0.01526&$^{86}_{41}Nb$&21.04950&0.00763&0.51880&0.00763\\
$^{87}_{40}Zr$&20.54596&0.01526&$^{87}_{41}Nb$&21.04950&0.00000&0.50354&0.01526\\
$^{88}_{40}Zr$&20.53833&0.00763&$^{88}_{41}Nb$&21.04950&0.00000&0.51117&0.00763 \\
$^{89}_{40}Zr$&20.53070&0.00763&$^{89}_{41}Nb$&21.05713&0.00763&0.52643&0.01526\\
$^{90}_{40}Zr$&20.53833&0.00763&$^{90}_{41}Nb$&21.05713&0.00000&0.51880&0.00764\\
$^{91}_{40}Zr$&20.53071&0.00762&$^{91}_{41}Nb$&21.04950&0.00763&0.51879&0.00001\\
$^{92}_{40}Zr$&20.53833&0.00762&$^{92}_{41}Nb$&21.05713&0.00763&0.51880&0.00001\\
$^{93}_{40}Zr$&20.53833&0.00000&$^{93}_{41}Nb$&21.05713&0.00000&0.51880&0.00000\\
$^{94}_{40}Zr$&20.53833&0.00000&$^{94}_{41}Nb$&21.05713&0.00000&0.51117&0.00763\\
$^{95}_{40}Zr$&20.53833&0.00000&$^{95}_{41}Nb$&21.04950&0.00763&0.51880&0.00763\\
$^{96}_{40}Zr$&20.53070&0.00763&$^{96}_{41}Nb$&21.04950&0.00000&0.51880&0.00000\\
$^{97}_{40}Zr$&20.53833&0.00763&$^{97}_{41}Nb$&21.05713&0.00763&0.51880&0.00000\\
$^{98}_{40}Zr$&20.53833&0.00000&$^{98}_{41}Nb$&21.05713&0.00000&0.51880&0.00000\\
\hline
\end{tabular}
\end{center}

\begin{center}
Table 1 (continued)    \\
\vspace*{0.2cm}
\begin{tabular}{||c|c|c|c|c|c|c|c||}
\hline Nuclei & $\Delta M$ &$\Delta$& Nuclei&$\Delta $&$\Delta$&$\delta$&$\delta\delta$ \\
\hline
$^{199}_{82}Pb$&42.43470&       &$^{199}_{83}Bi$&42.96875&     &0.53405&\\
$^{200}_{82}Pb$&42.43469&0.00001&$^{200}_{83}Bi$&42.96875&0.00000&0.53406&0.00001\\
$^{201}_{82}Pb$&42.43469&0.00000&$^{201}_{83}Bi$&42.95349&0.01526&0.51880&0.01526\\
$^{202}_{82}Pb$&42.44995&0.01526&$^{202}_{83}Bi$&42.98401&0.03052&0.53406&0.01526 \\
$^{203}_{82}Pb$&42.43470&0.01525&$^{203}_{83}Bi$&42.96875&0.01526&0.53405&0.00001\\
$^{204}_{82}Pb$&42.43469&0.00001&$^{204}_{83}Bi$&42.96875&0.00000&0.53406&0.00001\\
$^{205}_{82}Pb$&42.44996&0.01526&$^{205}_{83}Bi$&42.96875&0.00000&0.51879&0.01527\\
$^{206}_{82}Pb$&42.43469&0.01527&$^{206}_{83}Bi$&42.96875&0.00000&0.53406&0.01527\\
$^{207}_{82}Pb$&42.43469&0.00000&$^{207}_{83}Bi$&42.96875&0.00000&0.53406&0.00000\\
$^{208}_{82}Pb$&42.43469&0.00000&$^{208}_{83}Bi$&42.96875&0.00000&0.53406&0.00000\\
$^{209}_{82}Pb$&42.43469&0.00000&$^{209}_{83}Bi$&42.96875&0.00000&0.53406&0.00000\\
$^{210}_{82}Pb$&42.44995&0.01526&$^{210}_{83}Bi$&42.96875&0.00000&0.51880&0.01526\\
$^{211}_{82}Pb$&42.43469&0.01526&$^{211}_{83}Bi$&42.96875&0.00000&0.53406&0.01526\\
$^{212}_{82}Pb$&42.44995&0.01526&$^{212}_{83}Bi$&42.96875&0.00000&0.51880&0.01526\\
\hline
\end{tabular}
\end{center}

\begin{center}
Table 1 (continued)    \\
\vspace*{0.2cm}
\begin{tabular}{||c|c|c|c|c|c|c|c||}
\hline Nuclei & $\Delta M$ &$\Delta\Delta M$& Nuclei&$\Delta M$&$\Delta\Delta M$&$\delta M$&$\delta\delta M$ \\
\hline
$^{229}_{92}U$&47.72949&       &$^{229}_{93}Np$&48.26354&       &0.53405&\\
$^{230}_{92}U$&47.71423&0.01526&$^{230}_{93}Np$&48.24830&0.01524&0.53407&0.00002\\
$^{231}_{92}U$&47.71423&0.00000&$^{231}_{93}Np$&48.24829&0.00001&0.53406&0.00001\\
$^{232}_{92}U$&47.71424&0.00001&$^{233}_{93}Np$&48.24829&0.00000&0.53405&0.00001 \\
$^{233}_{92}U$&47.72951&0.01527&$^{234}_{93}Np$&48.24829&0.00000&0.51878&0.01527\\
$^{234}_{92}U$&47.71424&0.01527&$^{235}_{93}Np$&48.26355&0.01526&0.54931&0.03053\\
$^{235}_{92}U$&47.71423&0.00001&$^{236}_{93}Np$&48.23304&0.03051&0.51881&0.03050\\
$^{236}_{92}U$&47.72949&0.01526&$^{237}_{93}Np$&48.24829&0.01525&0.51880&0.00001\\
$^{237}_{92}U$&47.72949&0.00000&$^{238}_{93}Np$&48.24829&0.00000&0.51880&0.00000\\
$^{238}_{92}U$&47.71424&0.01525&$^{239}_{93}Np$&48.24829&0.00000&0.53405&0.01525\\
$^{239}_{92}U$&47.71423&0.00001&$^{240}_{93}Np$&48.24829&0.00000&0.53406&0.00000\\
$^{240}_{92}U$&47.71424&0.00001&$^{241}_{93}Np$&48.24829&0.00000&0.53405&0.00001\\
\hline
\end{tabular}
\end{center}

We demonstrated only a small part of our results in Tables 1. The
differences between nuclear and atomic rest masses are quantized:
$\Delta\Delta M=0.06,\;\;0.11-0.12,\;\;0.24, \;\;0.47-0.48,\;\;
0.94-0.96,\;\; 1.89-1.91, \;\;2.86,\;3.81-3.82 ,\;7.62-7.64,
\;\;15.25-15.27,\;\; 30.50-30.53\;keV/c^{2}$. The minimal value of
the quanta is equal to $\Delta\Delta M\approx 0.06\;keV/c^{2}$.

$\bullet$ The differences of nuclear and atomic rest masses are
quantized
$$\Delta\Delta M=M_{0}*2^{n},\;\;n=1,2,3,...\eqno(35)$$
where $M_{0}\approx 0.06\;keV/c^{2}$. It is essential to note that
this new phenomenological quantization rule of mass differences
introduced a simple doubling process of the mass quanta $M_{0}$.
What is a mechanism of such quantization?

\subsection{Electron Capture and $\beta$-decay}

The main quantity which characterizes  electron capture and
$\beta$-decay processes in nuclei is the rest mass differences
$\Delta M(A,Z,Z'=Z\pm 1,Z\pm 2)$ of the initial and final atoms
with the same atomic number.

We observed accidentally long time ago (seventies) that  some
quantities $\Delta M$ are the same or their ratios are equal to an
integer number for the atoms independent of $A$ and $Z$:
$$\Delta M=n*0.167847,\; n=1,2,3,...12$$
One can see from Table 2 that the quantities $\delta=\mid \Delta
M(exp)-n*0.167847 \mid$ are equal to zero or 0.00001 $MeV/c^{2}$
within the experimental errors.
\begin{center}
Table 2.    \\
The differences between atomic rest masses $\Delta M(exp)=M(A,Z)-M(A,Z'),Z-Z'=\pm 1,\pm 2$,
$\delta=\mid \Delta M(exp)-n*0.167847 \mid$.\\

\vspace*{0.2cm}
\begin{tabular}{||c|c|c|c|c|c||}
\hline Atoms & $\Delta M(exp)$ &$\Delta M=n*0.167847$& $\delta$  \\
\hline
$m(^{35}_{16}S)-m(^{35}_{17}Cl)$&0.16785&1*0.167847=0.167847&0.00000        \\
$m(^{96}_{40}Zr)-m(^{96}_{41}Nb)$&0.16785&1*0.167847=0.167847&0.00000\\
$m(^{100}_{43}Tc)-m(^{100}_{42}Mo)$&0.16785&1*0.167847=0.167847&0.00000\\
$m(^{111}_{47}Ag)-m(^{111}_{49}In)$&0.16784&1*0.167847=0.167847&0.00001 \\
$m(^{145}_{61}Pm)-m(^{145}_{60}Nd)$&0.16785&1*0.167847=0.167847&0.00000\\
$m(^{198}_{79}Au)-m(^{198}_{78}Pt)$&0.33569&2*0.167847=0.33569&0.00000\\
$m(^{153}_{64}Gd)-m(^{153}_{63}Eu)$&0.50354&3*0.167847=0.50354&0.00000\\
$m(^{177}_{71}Lu)-m(^{177}_{72}Hf)$&0.50354&3*0.167847=0.50354&0.00000\\
$m(^{183}_{73}Ta)-m(^{183}_{75}Re)$&0.50354&3*0.167847=0.50354&0.00000  \\
$m(^{127}_{54}Xe)-m(^{127}_{55}I)$&0.67139&4*0.167847=0.67139&0.00000 \\
$m(^{175}_{72}Hf)-m(^{175}_{71}Lu)$&0.67138&4*0.167847=0.67139&0.00001\\
$m(^{177}_{73}Ta)-m(^{177}_{71}Lu)$&0.67138&4*0.167847=0.67139&0.00001 \\
$m(^{70}_{30}Zn)-m(^{70}_{32}Ge)$&1.00708&6*0.167847=1.00708&0.00000\\
$m(^{129}_{55}Cs)-m(^{129}_{53}I)$&1.00708&6*0.167847=1.00708&0.00000\\
$m(^{185}_{76}Os)-m(^{185}_{75}Re)$&1.00708&6*0.167847=1.00708&0.00000  \\
$m(^{189}_{75}Re)-m(^{189}_{76}Os)$&1.00708&6*0.167847=1.00708&0.00000 \\
$m(^{75}_{32}Ge)-m(^{75}_{33}As)$&1.17493&7*0.167847=1.17493&0.00000\\
$m(^{102}_{46}Pd)-m(^{102}_{44}Ru)$&1.17492&7*0.167847=1.17493&0.00001 \\
$m(^{166}_{69}Tm)-m(^{166}_{67}Ho)$&1.17493&7*0.167847=1.17493&0.00000\\
$m(^{177}_{73}Ta)-m(^{177}_{72}Hf)$&1.17492&7*0.167847=1.17493&0.00001\\
$m(^{232}_{92}U)-m(^{232}_{91}Pa)$&1.34277&8*0.167847=1.34277&0.00000  \\
$m(^{244}_{94}Pu)-m(^{244}_{96}Cm)$&1.34277&8*0.167847=1.34277&0.00000 \\
$m(^{147}_{63}Eu)-m(^{147}_{61}Pm)$&1.51062&9*0.167847=1.51062&0.00000\\
$m(^{187}_{77}Ir)-m(^{187}_{76}Os)$&1.51062&9*0.167847=1.51062&0.00000 \\
$m(^{187}_{77}Ir)-m(^{187}_{75}Re)$&1.51062&9*0.167847=1.51062&0.00000  \\
$m(^{98}_{43}Tc)-m(^{98}_{42}Mo)$&1.67847&10*0.167847=1.67847&0.00000 \\
$m(^{168}_{69}Tm)-m(^{168}_{68}Er)$&1.67846&10*0.167847=1.67847&0.00001\\
$m(^{162}_{68}Er)-m(^{162}_{66}Dy)$&1.84631&11*0.167847=1.84632&0.00001 \\
$m(^{92}_{41}Nb)-m(^{92}_{40}Zr)$&2.01416&12*0.167847=2.01416&0.00000\\
$m(^{156}_{66}Dy)-m(^{156}_{64}Gd)$&2.01416&12*0.167847=2.01416&0.00000 \\

 \hline
\end{tabular}
\end{center}

We decided to investigate in a systematic way the differences
between atomic rest masses with the same atomic numbers in which
the number of electrons differs to one or two. The unique
restriction was the requirement that the final atoms (nuclei)
should be stable. We are not able to present all results only a
very small amount  shown below in Tables 3. The calculations were
performed by formula $\Delta M=n*0.0076294,\;MeV/c^{2},\;
n$-integer number.

\newpage
\begin{center}
Table 3.    \\
The differences between atomic rest masses $\Delta M,\;in\; MeV/c^{2}$\\
\vspace*{0.2cm}
\begin{tabular}{||c|c|c|c|c|c||}
\hline Atoms & $\Delta M$ &$\Delta=n*0.0076294\equiv n*a$& $\delta$  \\
\hline
$m(^{127}_{52}Tl)-m(^{127}_{54}Xe)$&0.03051&4*a=0.03052&0.00001 \\
$m(^{164}_{68}Er)-m(^{164}_{66}Dy)$&0.03052&4*a=0.03052&0.00000\\
$m(^{184}_{76}Os)-m(^{184}_{75}Re)$&0.04578&6*a=0.04578&0.00000 \\
$m(^{193}_{78}Pt)-m(^{193}_{77}Ir)$&0.06103&8*a=0.06103&0.00000 \\
$m(^{151}_{62}Sm)-m(^{151}_{63}Eu)$&0.06103&8*a=0.06103&0.00000\\
$m(^{136}_{54}Xe)-m(^{136}_{55}Cs)$&0.06866&9*a=0.06866&0.00000\\
$m(^{136}_{54}Xe)-m(^{136}_{58}Ce)$&0.06866&9*a=0.06866&0.00000 \\
$m(^{157}_{65}Tb)-m(^{157}_{64}Gd)$&0.07630&10*a=0.07629&0.00001 \\
$m(^{150}_{61}Pm)-m(^{150}_{60}Nd)$&0.09155&12*a=0.09155&0.00000\\
$m(^{244}_{95}Am)-m(^{244}_{94}Pu)$&0.09155&12*a=0.09155&0.00000\\
$m(^{130}_{56}Ba)-m(^{130}_{52}Te)$&0.09155&12*a=0.09155&0.00000\\
$m(^{171}_{69}Tm)-m(^{171}_{70}Yb)$&0.10681&14*a=0.10681&0.00000\\
$m(^{176}_{71}Lu)-m(^{176}_{70}Yb)$&0.10681&14*a=0.10681&0.00000  \\
$m(^{179}_{73}Ta)-m(^{179}_{72}Hf)$&0.10681&14*a=0.10681&0.00000 \\
$m(^{98}_{42}Mo)-m(^{98}_{44}Ru)$&0.11444&15*a=0.11444&0.00000\\
$m(^{193}_{76}Os)-m(^{193}_{77}Ir)$&0.11444&15*a=0.11444&0.00000 \\
$m(^{235}_{93}Np)-m(^{235}_{92}U)$&0.12207&16*a=0.12207&0.00000\\
$m(^{180}_{74}W)-m(^{180}_{72}Hf)$&0.13733&18*a=0.13733&0.00000 \\
$m(^{197}_{78}Pt)-m(^{197}_{80}Hg)$&0.13733&18*a=0.13733&0.00000\\
$m(^{238}_{93}Np)-m(^{238}_{92}U)$&0.15258&20*a=0.15259&0.00001 \\
$m(^{35}_{16}S)-m(^{35}_{17}Cl)$&0.16785&22*a=0.16785&0.00000        \\
$m(^{96}_{40}Zr)-m(^{96}_{41}Nb)$&0.16785&22*a=0.16785&0.00000\\
$m(^{100}_{43}Tc)-m(^{100}_{42}Mo)$&0.16785&22*a=0.16785&0.00000\\
$m(^{111}_{47}Ag)-m(^{111}_{49}In)$&0.16784&22*a=0.16785&0.00001 \\
$m(^{145}_{61}Pm)-m(^{145}_{60}Nd)$&0.16785&22*a=0.16785&0.00000\\
$m(^{178}_{71}Lu)-m(^{178}_{73}Ta)$&0.18310&24*a=0.18311&0.00001\\
$m(^{181}_{74}W)-m(^{181}_{73}Ta)$&0.18310&24*a=0.18311&0.00001\\
$m(^{106}_{46}Ag)-m(^{106}_{48}Gd)$&0.19836&26*a=0.19836&0.00000  \\
$m(^{190}_{77}Ir)-m(^{190}_{76}Os)$&0.19834&26*a=0.19836&0.00002 \\
$m(^{200}_{81}Tl)-m(^{200}_{79}Au)$&0.21363&28*a=0.21362&0.00001\\
$m(^{195}_{79}Au)-m(^{195}_{78}Pt)$&0.21363&28*a=0.21362&0.00001 \\
$m(^{175}_{72}Hf)-m(^{175}_{70}Yb)$&0.21362&28*a=0.21362&0.00000  \\
$m(^{103}_{44}Ru)-m(^{103}_{46}Pd)$&0.22126&29*a=0.22125&0.00001 \\
$m(^{147}_{61}Pm)-m(^{147}_{62}Sm)$&0.22888&30*a=0.22888&0.00000\\
$m(^{155}_{63}Eu)-m(^{155}_{64}Gd)$&0.22888&30*a=0.22888&0.00000 \\
$m(^{55}_{26}Fe)-m(^{55}_{25}Mn)$&0.22888&30*a=0.22888&0.00000\\
$m(^{71}_{32}Ge)-m(^{71}_{31}Ga)$&0.23651&31*a=0.23651&0.00000 \\
\hline
\end{tabular}
\end{center}

\newpage
\begin{center}
Table 3 (continued).    \\
\vspace*{0.2cm}
\begin{tabular}{||c|c|c|c|c|c||}
\hline
$m(^{194}_{79}Au)-m(^{194}_{77}Ir)$&0.24414&32*a=0.24414&0.00000 \\
$m(^{96}_{43}Tc)-m(^{96}_{44}Ru)$&0.25177&33*a=0.25177&0.00000\\
$m(^{168}_{69}Tm)-m(^{168}_{70}Yb)$&0.25940&34*a=0.25940&0.00000\\
$m(^{167}_{67}Ho)-m(^{167}_{69}Tm)$&0.25940&34*a=0.25940&0.00000 \\
$m(^{161}_{67}Ho)-m(^{161}_{65}Tb)$&0.25940&34*a=0.25940&0.00000\\
$m(^{108}_{48}Cd)-m(^{108}_{46}Pd)$&0.26703&35*a=0.26703&0.00000\\
$m(^{135}_{55}Cs)-m(^{135}_{56}Ba)$&0.26703&35*a=0.26703&0.00000\\
$m(^{158}_{66}Dy)-m(^{158}_{64}Gd)$&0.27465&36*a=0.27466&0.00001\\
$m(^{139}_{58}Ce)-m(^{139}_{57}La)$&0.27466&36*a=0.27466&0.00000  \\
$m(^{81}_{36}Kr)-m(^{81}_{35}Br)$&0.28229&37*a=0.28229&0.00000 \\
$m(^{174}_{71}Lu)-m(^{174}_{72}Hf)$&0.28992&38*a=0.28992&0.00000\\
$m(^{124}_{53}I)-m(^{124}_{54}Xe)$&0.28992&38*a=0.28992&0.00000 \\
$m(^{91}_{39}Y)-m(^{91}_{41}Nb)$&0.28992&38*a=0.28992&0.00000\\
$m(^{99}_{43}Tc)-m(^{99}_{44}Ru)$&0.29755&39*a=0.29755&0.00000 \\
$m(^{117}_{51}Sb)-m(^{117}_{49}In)$&0.29754&39*a=0.29755&0.00001\\
$m(^{153}_{62}Sm)-m(^{153}_{64}Gd)$&0.30517&40*a=0.30518&0.00001 \\
$m(^{162}_{67}Ho)-m(^{162}_{68}Er)$&0.30518&40*a=0.30518&0.00000        \\
$m(^{191}_{76}Os)-m(^{191}_{77}Ir)$&0.30518&40*a=0.30518&0.00000\\
$m(^{93}_{42}Mo)-m(^{93}_{40}Zr)$&0.31280&41*a=0.31281&0.00001\\
$m(^{75}_{32}Ge)-m(^{75}_{34}Se)$&0.31281&41*a=0.31281&0.00000 \\
$m(^{170}_{69}Tm)-m(^{170}_{68}Er)$&0.32043&42*a=0.32043&0.00000\\
$m(^{113}_{48}Cd)-m(^{113}_{49}In)$&0.32043&42*a=0.32043&0.00000\\
$m(^{97}_{43}Tc)-m(^{97}_{42}Mo)$&0.32043&42*a=0.32043&0.00000\\
$m(^{198}_{79}Au)-m(^{198}_{78}Pt)$&0.33569&44*a=0.33569&0.00000  \\
$m(^{73}_{33}As)-m(^{73}_{32}Ge)$&0.34332&45*a=0.34332&0.00000 \\
$m(^{131}_{55}Cs)-m(^{131}_{54}Xe)$&0.35095&46*a=0.35095&0.00000\\
$m(^{169}_{68}Er)-m(^{169}_{69}Tm)$&0.35095&46*a=0.35095&0.00000 \\
$m(^{204}_{81}Tl)-m(^{204}_{80}Hg)$&0.35095&46*a=0.35095&0.00000  \\
$m(^{54}_{24}Cr)-m(^{54}_{26}Fe)$&0.35095&46*a=0.35095&0.00000 \\
$m(^{72}_{33}As)-m(^{72}_{31}Ga)$&0.35095&46*a=0.35095&0.00000\\
$m(^{159}_{66}Dy)-m(^{159}_{65}Tb)$&0.36621&48*a=0.36621&0.00000 \\
$m(^{140}_{57}La)-m(^{140}_{59}Pr)$&0.36621&48*a=0.36621&0.00000\\
$m(^{130}_{55}Cs)-m(^{130}_{56}Ba)$&0.36621&48*a=0.36621&0.00000 \\
\hline
\end{tabular}
\end{center}

\newpage
\begin{center}
Table 3 (continued).    \\
\vspace*{0.2cm}
\begin{tabular}{||c|c|c|c|c|c||}
\hline
$m(^{122}_{50}Sn)-m(^{122}_{52}Te)$&0.36621&48*a=0.36621&0.00000 \\
$m(^{92}_{41}Nb)-m(^{92}_{42}Mo)$&0.36621&48*a=0.36621&0.00000\\
$m(^{85}_{38}Sr)-m(^{85}_{36}Kr)$&0.37384&49*a=0.37384&0.00000\\
$m(^{165}_{68}Er)-m(^{165}_{67}Ho)$&0.38147&50*a=0.38147&0.00000 \\
$m(^{121}_{50}Sn)-m(^{111}_{51}Sb)$&0.38910&51*a=0.38910&0.00000\\
$m(^{149}_{61}Pm)-m(^{149}_{63}Eu)$&0.39673&52*a=0.39673&0.00000\\
$m(^{151}_{64}Gd)-m(^{151}_{62}Sm)$&0.39673&52*a=0.39673&0.00000\\
$m(^{43}_{21}Sc)-m(^{43}_{19}K)$&0.40436&53*a=0.40436&0.00000\\
$m(^{93}_{42}Mo)-m(^{93}_{41}Nb)$&0.40436&53*a=0.40436&0.00000  \\
$m(^{192}_{76}Os)-m(^{192}_{78}Pt)$&0.41199&54*a=0.41199&0.00000 \\
$m(^{41}_{20}Ca)-m(^{41}_{19}K)$&0.41962&55*a=0.41962&0.00000\\
$m(^{67}_{31}Ga)-m(^{67}_{29}Cu)$&0.42725&56*a=0.42725&0.00000 \\
$m(^{130}_{53}I)-m(^{130}_{52}Te)$&0.42725&56*a=0.42725&0.00000\\
$m(^{133}_{54}Xe)-m(^{133}_{55}Cs)$&0.42725&56*a=0.42725&0.00000 \\
$m(^{156}_{65}Tb)-m(^{156}_{66}Dy)$&0.42725&56*a=0.42725&0.00000\\
$m(^{185}_{74}W)-m(^{185}_{75}Re)$&0.42724&56*a=0.42725&0.00001 \\
$m(^{204}_{80}Hg)-m(^{204}_{82}Pb)$&0.42724&56*a=0.42725&0.00001        \\
$m(^{25}_{13}Al)-m(^{25}_{11}Na)$&0.44250&58*a=0.30518&0.00000\\
$m(^{123}_{52}Te)-m(^{123}_{51}Sb)$&0.45777&60*a=0.45776&0.00001\\
$m(^{151}_{64}Gd)-m(^{151}_{63}Eu)$&0.45776&60*a=0.45776&0.00000 \\
$m(^{175}_{70}Yb)-m(^{175}_{71}Lu)$&0.45776&60*a=0.45776&0.00000\\
$m(^{199}_{79}Au)-m(^{199}_{80}Hg)$&0.45776&60*a=0.45776&0.00000\\
$m(^{136}_{57}La)-m(^{136}_{58}Ce)$&0.46539&61*a=0.46539&0.00000\\
$m(^{186}_{74}W)-m(^{186}_{76}Os)$&0.47302&62*a=0.47302&0.00000  \\
$m(^{203}_{80}Hg)-m(^{203}_{81}Tl)$&0.47302&62*a=0.47302&0.00000 \\
$m(^{232}_{91}Pa)-m(^{132}_{90}Th)$&0.48828&64*a=0.48828&0.00000\\
$m(^{203}_{82}Pb)-m(^{203}_{80}Hg)$&0.48829&64*a=0.48828&0.00001 \\
$m(^{201}_{81}Tl)-m(^{201}_{80}Hg)$&0.48828&64*a=0.48828&0.00000  \\
$m(^{189}_{77}Ir)-m(^{189}_{75}Re)$&0.48828&64*a=0.48828&0.00000 \\
$m(^{115}_{49}In)-m(^{115}_{50}Sn)$&0.49591&65*a=0.49591&0.00000\\
$m(^{153}_{64}Gd)-m(^{153}_{63}Eu)$&0.50354&66*a=0.50354&0.00000 \\
$m(^{177}_{71}Lu)-m(^{177}_{72}Hf)$&0.50354&66*a=0.50354&0.00000\\
$m(^{183}_{73}Ta)-m(^{183}_{75}Re)$&0.50354&66*a=0.50354&0.00000 \\
\hline
\end{tabular}
\end{center}

\newpage
\begin{center}
Table 3 (continued).    \\
\vspace*{0.2cm}
\begin{tabular}{||c|c|c|c|c|c||}
\hline Atoms & $\Delta M$ &$\Delta=n*0.0076294\equiv n*a$& $\delta$  \\
\hline
$m(^{86}_{37}Rb)-m(^{86}_{36}Te)$&0.51117&67*a=0.51117&0.00000 \\
$m(^{189}_{77}Ir)-m(^{189}_{76}Os)$&0.51880&68*a=0.51880&0.00000\\
$m(^{133}_{56}Ba)-m(^{133}_{55}Cs)$&0.52643&69*a=0.52643&0.00000\\
$m(^{114}_{48}Cd)-m(^{114}_{50}Sn)$&0.54168&71*a=0.54169&0.00001 \\
$m(^{144}_{61}Pm)-m(^{144}_{62}Sm)$&0.54931&72*a=0.54932&0.00001\\
$m(^{148}_{60}Nd)-m(^{148}_{61}Pm)$&0.54932&72*a=0.54932&0.00000\\
$m(^{101}_{45}Rh)-m(^{101}_{44}Ru)$&0.54932&72*a=0.54932&0.00000\\
$m(^{103}_{46}Pd)-m(^{103}_{45}Rh)$&0.54932&72*a=0.54932&0.00000\\
$m(^{10}_{5}B)-m(^{10}_{4}Be)$&0.55695&73*a=0.55695&0.00000  \\
$m(^{39}_{18}Ar)-m(^{39}_{19}K)$&0.56458&74*a=0.56458&0.00000 \\
$m(^{105}_{45}Rh)-m(^{105}_{46}Pd)$&0.56458&74*a=0.56458&0.00000\\
$m(^{169}_{70}Yb)-m(^{169}_{68}Er)$&0.56458&74*a=0.56458&0.00000 \\
$m(^{64}_{29}Cu)-m(^{64}_{30}Zn)$&0.57983&76*a=0.57983&0.00000\\
$m(^{125}_{51}Sb)-m(^{125}_{53}I)$&0.57983&76*a=0.57983&0.00000 \\
$m(^{137}_{55}Cs)-m(^{137}_{57}La)$&0.57984&76*a=0.57983&0.00001\\
$m(^{141}_{58}Ce)-m(^{141}_{59}Pr)$&0.57983&76*a=0.57983&0.00000 \\
$m(^{185}_{76}Os)-m(^{185}_{74}W)$&0.57984&76*a=0.57983&0.00001 \\
$m(^{124}_{54}Xe)-m(^{124}_{50}Sn)$&0.58747&77*a=0.58747&0.00000\\
$m(^{197}_{80}Hg)-m(^{197}_{79}Au)$&0.59510&78*a=0.59509&0.00001\\
$m(^{186}_{75}Re)-m(^{186}_{74}W)$&0.59509&78*a=0.59509&0.00000 \\
$m(^{161}_{65}Tb)-m(^{161}_{66}Dy)$&0.59510&78*a=0.59509&0.00001\\
$m(^{155}_{65}Tb)-m(^{155}_{63}Eu)$&0.59509&78*a=0.59509&0.00000\\
$m(^{53}_{25}Mn)-m(^{53}_{24}Cr)$&0.59510&78*a=0.59509&0.00001\\
$m(^{47}_{21}Sc)-m(^{47}_{22}Ti)$&0.60272&79*a=0.60272&0.00000  \\
$m(^{49}_{23}V)-m(^{49}_{22}Ti)$&0.60273&79*a=0.60272&0.00001 \\
$m(^{137}_{57}La)-m(^{137}_{56}Ba)$&0.60272&79*a=0.60272&0.00000\\
$m(^{159}_{64}Gd)-m(^{159}_{66}Dy)$&0.61035&80*a=0.61035&0.00000 \\
$m(^{190}_{77}Ir)-m(^{190}_{78}Pt)$&0.61035&80*a=0.61035&0.00000  \\
$m(^{124}_{51}Sb)-m(^{124}_{50}Sn)$&0.61798&81*a=0.61798&0.00000 \\
$m(^{131}_{53}I)-m(^{131}_{55}Cs)$&0.62561&82*a=0.62561&0.00000\\
$m(^{173}_{69}Tm)-m(^{173}_{71}Lu)$&0.62561&82*a=0.62561&0.00000 \\
$m(^{96}_{40}Zr)-m(^{96}_{44}Ru)$&0.63324&83*a=0.63324&0.00000\\
$m(^{172}_{71}Lu)-m(^{172}_{69}Tm)$&0.64087&84*a=0.64087&0.00000 \\
\hline
\end{tabular}
\end{center}

\newpage
\begin{center}
Table 3 (continued).    \\
\vspace*{0.2cm}
\begin{tabular}{||c|c|c|c|c|c||}
\hline Atoms & $\Delta M$ &$\Delta=n*0.0076294\equiv n*a$& $\delta$  \\
\hline
$m(^{121}_{52}Te)-m(^{121}_{50}Sn)$&0.64850&85*a=0.64850&0.00000 \\
$m(^{70}_{31}Ga)-m(^{70}_{30}Zn)$&0.64850&85*a=0.64850&0.00000\\
$m(^{170}_{68}Er)-m(^{170}_{70}Yb)$&0.65613&86*a=0.65613&0.00000\\
$m(^{209}_{82}Pb)-m(^{209}_{83}Bi)$&0.65613&86*a=0.65613&0.00000 \\
$m(^{112}_{49}In)-m(^{112}_{50}Sn)$&0.66376&87*a=0.66376&0.00000\\
$m(^{127}_{54}Xe)-m(^{127}_{53}I)$&0.67139&88*a=0.67139&0.00000\\
$m(^{175}_{72}Hf)-m(^{175}_{71}Lu)$&0.67138&88*a=0.67139&0.00001\\
$m(^{177}_{73}Ta)-m(^{177}_{71}Rh)$&0.67138&88*a=0.67139&0.00001\\
$m(^{196}_{79}Au)-m(^{196}_{80}Hg)$&0.67139&88*a=0.67139&0.00000  \\
$m(^{54}_{26}Fe)-m(^{54}_{24}Cr)$&0.67902&89*a=0.67902&0.00000 \\
$m(^{77}_{35}Br)-m(^{77}_{33}As)$&0.67902&89*a=0.67902&0.00000\\
$m(^{138}_{58}Ce)-m(^{138}_{56}Ba)$&0.68664&90*a=0.68665&0.00001 \\
$m(^{149}_{63}Eu)-m(^{149}_{62}Sm)$&0.68665&90*a=0.68665&0.00000\\
$m(^{173}_{71}Lu)-m(^{173}_{70}Yb)$&0.68665&90*a=0.68665&0.00000 \\
$m(^{188}_{77}Ir)-m(^{188}_{75}Re)$&0.68665&90*a=0.68665&0.00000\\
$m(^{85}_{36}Kr)-m(^{85}_{37}Rb)$&0.69428&91*a=0.69428&0.00000 \\
$m(^{127}_{52}Te)-m(^{127}_{53}I)$&0.70190&92*a=0.70190&0.00000 \\
$m(^{154}_{63}Eu)-m(^{154}_{62}Sm)$&0.70191&92*a=0.70190&0.00001\\
$m(^{54}_{25}Mn)-m(^{54}_{26}Fe)$&0.70190&92*a=0.70190&0.00000\\
$m(^{78}_{35}Br)-m(^{78}_{36}Kr)$&0.70190&92*a=0.70190&0.00000 \\
$m(^{180}_{73}Ta)-m(^{180}_{74}W)$&0.71716&94*a=0.71716&0.00000\\
$m(^{191}_{78}Pt)-m(^{191}_{76}Os)$&0.71717&94*a=0.71716&0.00001\\
$m(^{113}_{50}Sn)-m(^{113}_{48}Cd)$&0.71716&94*a=0.71716&0.00000\\
$m(^{167}_{67}Ho)-m(^{167}_{68}Er)$&0.73242&96*a=0.73242&0.00000  \\
$m(^{197}_{78}Pt)-m(^{197}_{79}Au)$&0.73242&96*a=0.73242&0.00000 \\
$m(^{52}_{25}Mn)-m(^{52}_{23}V)$&0.74005&97*a=0.74005&0.00000\\
$m(^{142}_{59}Pr)-m(^{142}_{58}Ce)$&0.74768&98*a=0.74768&0.00000 \\
$m(^{125}_{51}Sb)-m(^{125}_{52}Te)$&0.76294&100*a=0.76294&0.00000  \\
$m(^{95}_{43}Tc)-m(^{95}_{41}Nb)$&0.76294&100*a=0.76294&0.00000 \\
$m(^{201}_{79}Au)-m(^{201}_{81}Tl)$&0.76294&100*a=0.76294&0.00000\\
$m(^{103}_{44}Ru)-m(^{103}_{45}Rh)$&0.77057&101*a=0.77057&0.00000 \\
$m(^{105}_{47}Ag)-m(^{105}_{45}Rh)$&0.77819&102*a=0.77820&0.00001\\
$m(^{204}_{81}Tl)-m(^{204}_{80}Hg)$&0.77820&102*a=0.64087&0.00000 \\
\hline
\end{tabular}
\end{center}

\newpage
\begin{center}
Table 3 (continued).    \\
\vspace*{0.2cm}
\begin{tabular}{||c|c|c|c|c|c||}
\hline Atoms & $\Delta M$ &$\Delta=n*0.0076294\equiv n*a$& $\delta$  \\
\hline
$m(^{106}_{62}Sm)-m(^{106}_{63}Eu)$&0.80871&106*a=0.80871&0.00000 \\
$m(^{37}_{18}Ar)-m(^{37}_{17}Cl)$&0.81635&107*a=0.81635&0.00000\\
$m(^{134}_{54}Xe)-m(^{134}_{56}Ba)$&0.82397&108*a=0.89397&0.00000\\
$m(^{155}_{65}Tb)-m(^{155}_{64}Gd)$&0.82397&108*a=0.89397&0.00000 \\
$m(^{196}_{80}Hg)-m(^{196}_{78}Pt)$&0.82397&108*a=0.89397&0.00000\\
$m(^{132}_{56}Ba)-m(^{132}_{54}Xe)$&0.83161&109*a=0.83160&0.00001\\
$m(^{181}_{72}Hf)-m(^{181}_{74}W)$&0.83924&110*a=0.83923&0.00001\\
$m(^{232}_{90}Th)-m(^{232}_{92}U)$&0.85449&112*a=0.85449&0.00000\\
$m(^{161}_{67}Ho)-m(^{161}_{66}Dy)$&0.85450&112*a=0.85449&0.00001  \\
$m(^{180}_{73}Ta)-m(^{180}_{72}Hf)$&0.85449&112*a=0.85449&0.00000 \\
$m(^{54}_{25}Mn)-m(^{54}_{24}Cr)$&0.86212&113*a=0.86212&0.00000\\
$m(^{75}_{34}Se)-m(^{75}_{33}As)$&0.86212&113*a=0.86212&0.00000 \\
$m(^{56}_{27}Co)-m(^{56}_{25}Mn)$&0.86975&114*a=0.86975&0.00000\\
$m(^{111}_{49}In)-m(^{111}_{48}Ag)$&0.86976&114*a=0.86975&0.00001 \\
$m(^{128}_{52}Te)-m(^{128}_{54}Xe)$&0.86976&114*a=0.86975&0.00001\\
$m(^{58}_{27}Co)-m(^{58}_{28}Ni)$&0.89264&117*a=0.89264&0.00000 \\
$m(^{84}_{37}Rb)-m(^{84}_{38}Sr)$&0.89264&117*a=0.89264&0.00000 \\
$m(^{110}_{47}Ag)-m(^{110}_{46}Pd)$&0.89264&117*a=0.89264&0.00000\\
$m(^{126}_{54}Xe)-m(^{126}_{52}Te)$&0.89264&117*a=0.89264&0.00000\\
$m(^{94}_{41}Nb)-m(^{94}_{40}Zr)$&0.90027&118*a=0.90027&0.00000 \\
$m(^{109}_{46}Pd)-m(^{109}_{48}Cd)$&0.90027&118*a=0.90027&0.00000\\
$m(^{195}_{77}Ir)-m(^{195}_{77}Au)$&0.90027&118*a=0.90027&0.00000\\
$m(^{69}_{30}Zn)-m(^{69}_{31}Ga)$&0.90790&119*a=0.90790&0.00000\\
$m(^{76}_{33}As)-m(^{76}_{32}Ge)$&0.92316&121*a=0.92316&0.00000  \\
$m(^{95}_{41}Nb)-m(^{95}_{42}Mo)$&0.93078&122*a=0.93079&0.00001 \\
$m(^{135}_{57}La)-m(^{135}_{55}Cs)$&0.93078&122*a=0.93079&0.00001\\
$m(^{143}_{59}Pr)-m(^{143}_{60}Nd)$&0.93079&112*a=0.93079&0.00000 \\
$m(^{158}_{65}Tb)-m(^{158}_{66}Dy)$&0.94605&124*a=0.94605&0.00000  \\
$m(^{164}_{67}Ho)-m(^{164}_{68}Er)$&0.96130&126*a=0.96130&0.00000 \\
$m(^{203}_{82}Pb)-m(^{203}_{81}Tl)$&0.96131&126*a=0.96130&0.00001\\
$m(^{159}_{64}Gd)-m(^{159}_{65}Tb)$&0.97656&128*a=0.97656&0.00000 \\
$m(^{131}_{53}I)-m(^{131}_{44}Xe)$&0.97656&128*a=0.97656&0.00000\\
$m(^{120}_{51}Sb)-m(^{120}_{52}Te)$&0.97656&128*a=0.97656&0.00000 \\
$m(^{170}_{69}Tm)-m(^{170}_{70}Yb)$&0.97656&128*a=0.97656&0.00000\\
\hline
\end{tabular}
\end{center}

\newpage
\begin{center}
Table  (continued).    \\
\vspace*{0.2cm}
\begin{tabular}{||c|c|c|c|c|c||}
\hline Atoms & $\Delta M$ &$\Delta=n*0.0076294\equiv n*a$& $\delta$  \\
\hline
$m(^{46}_{20}Ca)-m(^{46}_{22}Ti)$&0.99182&130*a=0.99182&0.00000 \\
$m(^{207}_{81}Tl)-m(^{207}_{83}Bi)$&0.99182&130*a=0.99182&0.00000\\
$m(^{199}_{81}Tl)-m(^{199}_{79}Au)$&0.99182&130*a=0.99182&0.00000\\
$m(^{182}_{75}Re)-m(^{182}_{73}Ta)$&0.99182&130*a=0.99182&0.00000 \\
$m(^{167}_{67}Ho)-m(^{167}_{68}Er)$&0.99182&130*a=0.99182&0.00000\\
$m(^{164}_{67}Ho)-m(^{164}_{66}Dy)$&0.99182&130*a=0.99182&0.00000\\
$m(^{38}_{19}K)-m(^{38}_{17}Cl)$&0.99946&131*a=0.99946&0.00000\\
$m(^{70}_{30}Zn)-m(^{70}_{32}Ge)$&1.00708&132*a=1.00708&0.00000\\
$m(^{129}_{55}Cs)-m(^{129}_{53}I)$&1.00708&132*a=1.00708&0.00000 \\
$m(^{185}_{76}Os)-m(^{185}_{75}Re)$&1.00708&132*a=1.00708&0.00000 \\
$m(^{189}_{75}Re)-m(^{189}_{76}Os)$&1.00708&132*a=1.00708&0.00000\\
$m(^{181}_{72}Hf)-m(^{181}_{73}Ta)$&1.02234&134*a=1.02234&0.00000 \\
$m(^{191}_{78}Pt)-m(^{191}_{76}Os)$&1.02234&134*a=1.02234&0.00000\\
$m(^{50}_{23}V)-m(^{50}_{24}Cr)$&1.03760&136*a=1.03760&0.00000 \\
$m(^{111}_{47}Ag)-m(^{111}_{48}Cd)$&1.03760&136*a=1.03760&0.00000\\
$m(^{113}_{50}Sn)-m(^{113}_{49}In)$&1.03759&136*a=1.03760&0.00001 \\
$m(^{121}_{52}Te)-m(^{121}_{51}Sb)$&1.03760&136*a=1.03760&0.00000 \\
$m(^{138}_{57}La)-m(^{138}_{58}Ce)$&1.03760&136*a=1.03760&0.00000\\
$m(^{143}_{61}Pm)-m(^{143}_{60}Nd)$&1.05286&138*a=1.05286&0.00000\\
$m(^{192}_{77}Ir)-m(^{192}_{76}Os)$&1.05285&138*a=1.05286&0.00001 \\
$m(^{198}_{78}Pt)-m(^{198}_{80}Hg)$&1.05285&138*a=1.05286&0.00001\\
$m(^{183}_{73}Ta)-m(^{183}_{74}W)$&1.06811&140*a=1.06812&0.00001\\
$m(^{186}_{75}Re)-m(^{186}_{76}Os)$&1.06811&140*a=1.06812&0.00001\\
$m(^{85}_{38}Sr)-m(^{85}_{37}Rb)$&1.06812&140*a=1.06812&0.00000  \\
$m(^{101}_{43}Tc)-m(^{101}_{45}Rh)$&1.07574&141*a=1.07575&0.00001 \\
$m(^{149}_{61}Pm)-m(^{149}_{62}Sm)$&1.08338&142*a=1.08337&0.00001\\
$m(^{176}_{70}Yb)-m(^{176}_{72}Hf)$&1.08337&142*a=1.08337&0.00000 \\
$m(^{193}_{76}Os)-m(^{193}_{78}Pt)$&1.08338&142*a=1.08338&0.00001  \\
$m(^{174}_{72}Hf)-m(^{174}_{70}Yb)$&1.09863&144*a=1.09863&0.00000 \\
$m(^{195}_{77}Ir)-m(^{195}_{78}Pt)$&1.11389&146*a=1.11389&0.00000\\
$m(^{109}_{46}Pd)-m(^{109}_{47}Ag)$&1.11389&146*a=1.11389&0.00000 \\
$m(^{238}_{92}U)-m(^{238}_{94}Pu)$&1.12916&148*a=1.12915&0.00001\\
$m(^{104}_{45}Rh)-m(^{104}_{44}Ru)$&1.13678&149*a=1.13678&0.00000 \\
$m(^{193}_{76}Os)-m(^{193}_{77}Ir)$&1.14441&150*a=1.14441&0.00000\\
\hline
\end{tabular}
\end{center}

\newpage
\begin{center}
Table 3 (continued).    \\
\vspace*{0.2cm}
\begin{tabular}{||c|c|c|c|c|c||}
\hline Atoms & $\Delta M$ &$\Delta=n*0.0076294\equiv n*a$& $\delta$  \\
\hline
$m(^{36}_{16}S)-m(^{36}_{17}Cl)$&1.14441&150*a=1.14441&0.00000 \\
$m(^{94}_{40}Zr)-m(^{94}_{42}Mo)$&1.14441&150*a=1.14441&0.00000\\
$m(^{102}_{45}Rh)-m(^{102}_{46}Pd)$&1.15204&151*a=1.15204&0.00000\\
$m(^{210}_{83}Bi)-m(^{210}_{84}Po)$&1.15966&152*a=1.15967&0.00001 \\
$m(^{75}_{32}Ge)-m(^{75}_{33}As)$&1.17493&154*a=1.17493&0.00000\\
$m(^{102}_{46}Pd)-m(^{102}_{44}Ru)$&1.17492&154*a=1.17493&0.00001\\
$m(^{166}_{69}Tm)-m(^{166}_{67}Ho)$&1.17493&154*a=1.17493&0.00000\\
$m(^{177}_{73}Ta)-m(^{177}_{72}Hf)$&1.17492&154*a=1.17493&0.00001\\
$m(^{137}_{55}Cs)-m(^{137}_{56}Ba)$&1.18256&155*a=1.18256&0.00000  \\
$m(^{176}_{71}Lu)-m(^{176}_{72}Hf)$&1.19018&156*a=1.19019&0.00001 \\
$m(^{129}_{55}Cs)-m(^{129}_{54}Xe)$&1.19018&156*a=1.19019&0.00000\\
$m(^{135}_{57}La)-m(^{135}_{56}Ba)$&1.19781&157*a=1.19782&0.00001 \\
$m(^{74}_{34}Se)-m(^{74}_{32}Ge)$&1.20545&158*a=1.20545&0.00000\\
$m(^{54}_{25}Mn)-m(^{54}_{26}Fe)$&1.21307&159*a=1.21307&0.00000 \\
$m(^{134}_{55}Cs)-m(^{134}_{56}Ba)$&1.22071&160*a=1.22070&0.00001\\
$m(^{158}_{65}Tb)-m(^{158}_{64}Gd)$&1.22070&160*a=1.22070&0.00000 \\
$m(^{141}_{60}Nd)-m(^{141}_{58}Ce)$&1.25122&164*a=1.25122&0.00000 \\
$m(^{154}_{62}Sm)-m(^{154}_{64}Gd)$&1.25122&164*a=1.25122&0.00000\\
$m(^{201}_{79}Au)-m(^{201}_{80}Hg)$&1.25122&164*a=1.25122&0.00000\\
$m(^{209}_{84}Po)-m(^{209}_{82}Pb)$&1.25122&164*a=1.25122&0.00000 \\
$m(^{73}_{31}Ga)-m(^{73}_{33}As)$&1.25122&164*a=1.25122&0.00000\\
$m(^{91}_{41}Nb)-m(^{91}_{40}Zr)$&1.25122&164*a=1.25122&0.00000\\
$m(^{128}_{53}I)-m(^{128}_{52}Te)$&1.25122&164*a=1.25122&0.00000\\
$m(^{86}_{36}Kr)-m(^{86}_{38}Sr)$&1.25885&165*a=1.25885&0.00000  \\
$m(^{126}_{53}I)-m(^{126}_{54}Xe)$&1.25885&165*a=1.25885&0.00000 \\
$m(^{132}_{55}Cs)-m(^{132}_{56}Ba)$&1.28173&168*a=1.28174&0.00001\\
$m(^{238}_{93}Np)-m(^{238}_{94}Pu)$&1.28173&168*a=1.28174&0.00001 \\
$m(^{157}_{63}Eu)-m(^{158}_{65}Tb)$&1.29699&170*a=1.29700&0.00001  \\
$m(^{81}_{34}Se)-m(^{81}_{36}Kr)$&1.30462&171*a=1.30463&0.00001 \\
$m(^{104}_{44}Ru)-m(^{104}_{46}Pd)$&1.30463&171*a=1.30463&0.00000\\
$m(^{173}_{69}Tm)-m(^{173}_{70}Yb)$&1.31226&172*a=1.31226&0.00000 \\
$m(^{40}_{19}K)-m(^{40}_{20}Ca)$&1.31226&172*a=1.31226&0.00000\\
$m(^{69}_{32}Ge)-m(^{69}_{30}Zn)$&1.31988&173*a=1.31988&0.00000 \\
$m(^{89}_{40}Zr)-m(^{89}_{38}Sr)$&1.33514&175*a=1.33514&0.00000\\
\hline
\end{tabular}
\end{center}

\newpage
\begin{center}
Table 3 (continued).    \\
\vspace*{0.2cm}
\begin{tabular}{||c|c|c|c|c|c||}
\hline Atoms & $\Delta M$ &$\Delta=n*0.0076294\equiv n*a$& $\delta$  \\
\hline
$m(^{105}_{47}Ag)-m(^{105}_{46}Pd)$&1.34277&176*a=1.34277&0.00000 \\
$m(^{232}_{91}Pa)-m(^{232}_{92}U)$&1.34277&176*a=1.34277&0.00000\\
$m(^{244}_{94}Pu)-m(^{244}_{96}Cm)$&1.34277&176*a=1.34277&0.00000\\
$m(^{65}_{30}Zn)-m(^{65}_{29}Cu)$&1.35040&177*a=1.35040&0.00000 \\
$m(^{74}_{33}As)-m(^{74}_{34}Se)$&1.35040&177*a=1.35040&0.00000\\
$m(^{123}_{50}Sn)-m(^{123}_{52}Te)$&1.35040&177*a=1.35040&0.00000\\
$m(^{77}_{35}Br)-m(^{77}_{34}Se)$&1.36566&177*a=1.36566&0.00000\\
$m(^{157}_{63}Eu)-m(^{157}_{64}Cd)$&1.37329&180*a=1.37329&0.00000\\
$m(^{171}_{71}Lu)-m(^{171}_{69}Tm)$&1.37329&180*a=1.37329&0.00000  \\
$m(^{190}_{78}Pt)-m(^{190}_{76}Os)$&1.37329&180*a=1.37329&0.00000 \\
$m(^{202}_{81}Tl)-m(^{202}_{80}Hg)$&1.37329&180*a=1.37329&0.00000\\
$m(^{54}_{25}Mn)-m(^{54}_{24}Cr)$&1.38092&181*a=1.38092&0.00000 \\
$m(^{107}_{48}Cd)-m(^{107}_{46}Pd)$&1.38855&182*a=1.38855&0.00000\\
$m(^{174}_{71}Lu)-m(^{174}_{70}Yb)$&1.38855&182*a=1.38855&0.00000 \\
$m(^{198}_{79}Au)-m(^{198}_{80}Hg)$&1.38855&182*a=1.38855&0.00000\\
$m(^{179}_{71}Lu)-m(^{179}_{72}Hf)$&1.40381&184*a=1.40381&0.00000 \\
$m(^{49}_{21}Sc)-m(^{49}_{23}V)$&1.40381&184*a=1.40381&0.00000 \\
$m(^{123}_{50}Sn)-m(^{123}_{51}Sb)$&1.41144&185*a=1.41144&0.00000\\
$m(^{142}_{58}Ce)-m(^{142}_{60}Nd)$&1.41907&186*a=1.41907&0.00000\\
$m(^{168}_{70}Yb)-m(^{168}_{68}Er)$&1.41906&186*a=1.41907&0.00001 \\
$m(^{207}_{81}Tl)-m(^{207}_{82}Pb)$&1.41907&186*a=1.41907&0.00000\\
$m(^{235}_{91}Pa)-m(^{235}_{92}U)$&1.41907&186*a=1.41907&0.00000\\
$m(^{244}_{95}Am)-m(^{244}_{96}Cm)$&1.43432&188*a=1.43432&0.00000\\
$m(^{114}_{49}In)-m(^{114}_{48}Cd)$&1.44959&190*a=1.44959&0.00000  \\
$m(^{184}_{76}Os)-m(^{184}_{74}W)$&1.44958&190*a=1.44959&0.00001 \\
$m(^{199}_{81}Tl)-m(^{199}_{80}Hg)$&1.44958&190*a=1.44959&0.00001\\
$m(^{117}_{49}In)-m(^{117}_{50}Sn)$&1.45722&191*a=1.45722&0.00000 \\
$m(^{146}_{61}Pm)-m(^{146}_{60}Nd)$&1.46484&192*a=1.46484&0.00000  \\
$m(^{192}_{77}Ir)-m(^{192}_{78}Pt)$&1.46484&192*a=1.46484&0.00000 \\
$m(^{79}_{36}Kr)-m(^{79}_{34}Se)$&1.46485&192*a=1.46484&0.00001\\
$m(^{184}_{75}Re)-m(^{184}_{74}W)$&1.49536&196*a=1.49536&0.00000 \\
$m(^{196}_{79}Au)-m(^{196}_{78}Pt)$&1.49536&196*a=1.49536&0.00000\\
$m(^{89}_{38}Sr)-m(^{89}_{39}Y)$&1.49536&196*a=1.49536&0.00000 \\
$m(^{40}_{18}Ar)-m(^{40}_{19}K)$&1.50299&197*a=1.50299&0.00000\\
\hline
\end{tabular}
\end{center}

\newpage
\begin{center}
Table 3 (continued).    \\
\vspace*{0.2cm}
\begin{tabular}{||c|c|c|c|c|c||}
\hline Atoms & $\Delta M$ &$\Delta=n*0.0076294\equiv n*a$& $\delta$  \\
\hline
$m(^{147}_{63}Eu)-m(^{147}_{61}Pm)$&1.51062&198*a=1.51062&0.00000 \\
$m(^{187}_{77}Ir)-m(^{187}_{76}Os)$&1.51062&198*a=1.51062&0.00000\\
$m(^{187}_{77}Ir)-m(^{187}_{75}Re)$&1.51062&198*a=1.51062&0.00000\\
$m(^{206}_{81}Tl)-m(^{206}_{82}Pb)$&1.52588&200*a=1.52588&0.00000 \\
$m(^{146}_{61}Pm)-m(^{146}_{62}Sm)$&1.52588&200*a=1.52588&0.00000\\
$m(^{205}_{80}Hg)-m(^{205}_{81}Tl)$&1.54114&202*a=1.54114&0.00000\\
$m(^{91}_{39}Y)-m(^{91}_{40}Zr)$&1.54113&202*a=1.54114&0.00001\\
$m(^{59}_{26}Fe)-m(^{59}_{27}Co)$&1.56403&205*a=1.56403&0.00000\\
$m(^{202}_{79}Au)-m(^{202}_{81}Tl)$&1.57165&206*a=1.57165&0.00000  \\
$m(^{81}_{34}Se)-m(^{81}_{35}Br)$&1.58691&208*a=1.58691&0.00000 \\
$m(^{87}_{39}Yb)-m(^{87}_{38}Rb)$&1.58691&208*a=1.58691&0.00000\\
$m(^{73}_{31}Ga)-m(^{73}_{32}Ge)$&1.59454&209*a=1.59454&0.00000 \\
$m(^{97}_{41}Nb)-m(^{97}_{43}Tc)$&1.61743&212*a=1.61743&0.00000\\
$m(^{122}_{51}Sb)-m(^{122}_{52}Te)$&1.61743&212*a=1.61743&0.00000 \\
$m(^{101}_{43}Tc)-m(^{101}_{44}Ru)$&1.62506&213*a=1.62506&0.00000\\
$m(^{79}_{36}Kr)-m(^{79}_{35}Br)$&1.62507&213*a=1.62506&0.00001 \\
$m(^{145}_{59}Pr)-m(^{145}_{61}Pm)$&1.63269&214*a=1.63269&0.00000 \\
$m(^{92}_{42}Mo)-m(^{92}_{40}Zr)$&1.64795&216*a=1.64795&0.00000\\
$m(^{108}_{47}Ag)-m(^{108}_{48}Cd)$&1.64795&216*a=1.64795&0.00000\\
$m(^{18}_{9}F)-m(^{18}_{8}O)$&1.65558&217*a=1.65558&0.00000 \\
$m(^{70}_{31}Ga)-m(^{70}_{32}Ge)$&1.65558&217*a=1.65558&0.00000\\
$m(^{95}_{43}Tc)-m(^{95}_{42}Mo)$&1.69372&222*a=1.69373&0.00001\\
$m(^{88}_{37}Rb)-m(^{88}_{39}Y)$&1.69372&222*a=1.69373&0.00000\\
$m(^{120}_{52}Te)-m(^{120}_{50}Sn)$&1.70135&223*a=1.70136&0.00001  \\
$m(^{138}_{57}La)-m(^{138}_{56}Ba)$&1.72424&226*a=1.72424&0.00000 \\
$m(^{147}_{53}Eu)-m(^{147}_{62}Sm)$&1.73950&228*a=1.73950&0.00000\\
$m(^{99}_{45}Rh)-m(^{99}_{43}Tc)$&1.74713&229*a=1.74713&0.00000 \\
$m(^{117}_{51}Sb)-m(^{117}_{50}Nd)$&1.75476&230*a=1.75476&0.00000  \\
$m(^{86}_{37}Rb)-m(^{86}_{38}Sr)$&1.77002&232*a=1.77002&0.00000 \\
$m(^{144}_{62}Sm)-m(^{144}_{60}Nd)$&1.78528&234*a=1.78528&0.00001\\
$m(^{84}_{38}Sr)-m(^{84}_{36}Kr)$&1.78528&234*a=1.78528&0.00000 \\
$m(^{98}_{43}Tc)-m(^{98}_{44}Ru)$&1.79291&235*a=1.79291&0.00000\\
$m(^{145}_{59}Pr)-m(^{145}_{60}Nd)$&1.80054&236*a=1.80054&0.00000 \\
$m(^{163}_{65}Tb)-m(^{163}_{66}Dy)$&1.80054&236*a=1.80054&0.00000\\
$m(^{163}_{65}Tb)-m(^{163}_{67}Ho)$&1.80054&236*a=1.80054&0.00000\\
\hline
\end{tabular}
\end{center}

\newpage
\begin{center}
Table 3 (continued).    \\
\vspace*{0.2cm}
\begin{tabular}{||c|c|c|c|c|c||}
\hline Atoms & $\Delta M$ &$\Delta=n*0.0076294\equiv n*a$& $\delta$  \\
\hline
$m(^{45}_{22}Ti)-m(^{45}_{20}Ca)$&1.80816&237*a=1.80817&0.00001 \\
$m(^{152}_{63}Eu)-m(^{152}_{64}Gd)$&1.81579&238*a=1.81580&0.00001\\
$m(^{182}_{73}Ta)-m(^{182}_{74}W)$&1.81580&238*a=1.81580&0.00000\\
$m(^{43}_{19}K)-m(^{43}_{20}Ca)$&1.81580&238*a=1.81580&0.00000 \\
$m(^{141}_{60}Nd)-m(^{141}_{59}Pr)$&1.83105&240*a=1.83106&0.00001\\
$m(^{162}_{68}Er)-m(^{162}_{66}Dy)$&1.84631&242*a=1.84631&0.00000\\
$m(^{166}_{67}Ho)-m(^{166}_{68}Er)$&1.86157&244*a=1.86157&0.00000\\
$m(^{87}_{39}Yb)-m(^{87}_{38}Sr)$&1.86157&244*a=1.86157&0.00000\\
$m(^{80}_{35}Br)-m(^{80}_{34}Se)$&1.87684&246*a=1.87683&0.00001  \\
$m(^{152}_{63}Eu)-m(^{152}_{62}Sm)$&1.87683&246*a=1.87683&0.00000 \\
$m(^{172}_{69}Tm)-m(^{172}_{70}Yb)$&1.89209&248*a=1.89209&0.00000\\
$m(^{209}_{84}Po)-m(^{209}_{83}Bi)$&1.90735&250*a=1.90735&0.00000 \\
$m(^{108}_{47}Ag)-m(^{108}_{46}Pd)$&1.91498&251*a=1.91498&0.00000\\
$m(^{112}_{50}Sn)-m(^{112}_{48}Cd)$&1.92260&252*a=1.92261&0.00001 \\
$m(^{148}_{60}Nd)-m(^{148}_{62}Sm)$&1.92261&252*a=1.92261&0.00000\\
$m(^{178}_{73}Ta)-m(^{178}_{71}Lu)$&1.92261&252*a=1.92261&0.00000 \\
$m(^{123}_{50}Sn)-m(^{123}_{51}Sb)$&1.93023&253*a=1.93024&0.00000 \\
$m(^{97}_{41}Nb)-m(^{97}_{42}Mo)$&1.93787&254*a=1.93787&0.00000\\
$m(^{154}_{63}Eu)-m(^{154}_{64}Gd)$&1.95313&256*a=1.95313&0.00000\\
$m(^{190}_{77}Ir)-m(^{190}_{76}Os)$&1.98364&260*a=1.98364&0.00000 \\
$m(^{122}_{51}Sb)-m(^{122}_{52}Te)$&1.98365&260*a=1.98364&0.00001\\
$m(^{114}_{49}In)-m(^{114}_{50}Sn)$&1.99127&261*a=1.99127&0.00000\\
$m(^{110}_{46}Pd)-m(^{110}_{48}Cd)$&1.99890&262*a=1.99890&0.00000\\
$m(^{49}_{21}Sc)-m(^{49}_{22}Ti)$&2.00654&263*a=2.00653&0.00001  \\
$m(^{80}_{35}Br)-m(^{80}_{36}Kr)$&2.00653&263*a=2.00653&0.00000 \\
$m(^{92}_{41}Nb)-m(^{92}_{40}Zr)$&2.01416&264*a=2.01416&0.00000\\
$m(^{156}_{66}Dy)-m(^{156}_{64}Gd)$&2.01416&264*a=2.01416&0.00000 \\
$m(^{139}_{56}Ba)-m(^{139}_{58}Ce)$&2.02942&266*a=2.02942&0.00000  \\
$m(^{94}_{41}Nb)-m(^{94}_{42}Mo)$&2.04468&268*a=2.04468&0.00000 \\
$m(^{99}_{45}Rh)-m(^{99}_{44}Ru)$&2.04468&268*a=2.04468&0.00000\\
$m(^{134}_{55}Cs)-m(^{134}_{56}Ba)$&2.04468&268*a=2.04468&0.00000 \\
$m(^{178}_{71}Lu)-m(^{178}_{72}Hf)$&2.10571&276*a=2.10571&0.00000\\
$m(^{132}_{55}Cs)-m(^{132}_{54}Xe)$&2.11334&277*a=2.11334&0.00000 \\
$m(^{188}_{75}Re)-m(^{188}_{76}Os)$&2.12097&278*a=2.12097&0.00000\\
$m(^{128}_{53}I)-m(^{128}_{54}Xe)$&1.12098&278*a=2.12097&0.00001\\
\hline
\end{tabular}
\end{center}
The rest mass differences of atoms  in the $\beta-decay$ (single
and double) and electron capture (single and double) processes are
quantized by the formula

$$M=\frac{n_{1}}{n_{2}}*0.0076294, \;(MeV/c^{2}), \eqno(35) $$
where $n_{1}$ are integer numbers and $n_{2}=1,2,4,8$. The cases
with $n_{2}=1$ and $n_{1}=4,...278$ are presented in Tables 3.
\newpage
\subsection{$\alpha$-decay}

The differences between atomic rest masses in the case of
$\alpha$-decay were calculated by the formula
$$\Delta M= m(^{A}_{Z}X)-m(^{A-4}_{Z-2}Y)-m(\alpha)$$
and are presented in Tables 4.
\begin{center}
Table 4.    \\
\vspace*{0.2cm}
\begin{tabular}{||c|c|c|c|c||}
\hline Nuclei & $\Delta M$ &$\Delta\Delta M$& n*0.0076294&$\delta $ \\
\hline
$^{237}_{93}Np$&4.96791&       &          &\\
$^{233}_{92}U$&4.90689&0.06102&8*a=0.06103&0.00001\\
$^{229}_{90}Th$&5.16630&0.25941&34*a=0.25940&0.00001\\
$^{225}_{89}Ac$&5.92923&0.76293&100*a=0.76294&0.00001 \\
$^{221}_{87}Fr$&6.44803&0.51880&68*a=0.51880&0.00000\\
$^{217}_{85}At$&7.22623&0.77820&102*a=0.77820&0.00000\\
$^{213}_{83}Bi$&5.97500&1.25123&164*a=1.25122&0.00001\\
$^{213}_{84}Po$&8.52323&2.54823&334*a=2.54822&0.00001\\
$^{228}_{90}Th$&5.51724&3.00599&394*a=3.00598&0.00001\\
$^{224}_{88}Ra$&5.79190&0.27466&36*a=0.27466&0.00000\\
$^{220}_{86}Rn$&6.40225&0.61035&80*a=0.61035&0.00000\\
$^{216}_{84}Po$&6.90580&0.50355&66*a=0.50354&0.00001\\
$^{212}_{83}Bi$&6.20389&0.70191&92*a=0.70190&0.00001\\
$^{212}_{84}Po$&8.93522&2.73133&358*a=2.73133&0.00000\\
$^{218}_{86}Rn$&7.25674&1.67848&220*a=1.67847&0.00001\\
$^{218}_{85}At$&6.87528&0.38146&50*a=0.38147&0.00001\\
$^{214}_{83}Bi$&5.62406&1.25122&164*a=1.25122&0.00000\\
\hline
\end{tabular}
\end{center}

\begin{center}
Table 4(continued).    \\
\vspace*{0.2cm}
\begin{tabular}{||c|c|c|c|c|c|c|c||}
\hline Nuclei & $\Delta M$ &$\Delta\Delta M$& n*0.0076294&$\delta$ \\
\hline
$^{235}_{92}U$&4.67800&        &       &\\
$^{231}_{91}Pa$&5.15103&0.47303&62*a=0.47302&0.00001\\
$^{227}_{89}Ac$&5.04422&0.10681&14*a=0.10681&0.00000\\
$^{223}_{87}Fr$&5.42569&0.38147&50*a=0.38147&0.00000 \\
$^{219}_{85}At$&6.38700&0.96131&126*a=0.96130&0.00001\\
$^{227}_{90}Th$&6.14286&0.24414&32*a=0.24414&0.00000\\
$^{223}_{88}Ra$&5.97500&0.16786&22*a=0.16785&0.00001\\
$^{219}_{86}Rn$&6.93631&0.96131&126*a=0.96130&0.00001\\
$^{215}_{84}Po$&7.53140&0.59509&78*a=0.59509&0.00000\\
$^{215}_{85}At$&8.17227&0.64087&84*a=0.64087&0.00000\\
$^{211}_{33}Bi$&6.76847&1.40380&184*a=1.40381&0.00001\\
$^{238}_{92}U$&4.26603&2.50244&328*a=2.50244&0.00000\\
$^{234}_{92}U$&4.86112&0.59509&78*a=0.59509&0.00000\\
$^{230}_{90}Th$&4.75430&0.10681&14*a=0.10681&0.00000\\
$^{226}_{88}Ra$&4.89164&0.13734&18*a=0.13733&0.00001\\
$^{222}_{86}Ru$&5.57828&0.68664&90*a=0.68665&0.00001\\
$^{218}_{84}Po$&6.11233&0.53405&70*a=0.53406&0.00001\\
$^{210}_{83}Bi$&5.04422&1.06811&140*a=1.06812&0.00001\\
$^{210}_{84}Po$&5.41044&0.36622&48*a=0.36621&0.00000\\
\hline
\end{tabular}
\end{center}

\section{Conclusion}

We come to the conclusion that  all atomic models  based on either
the Newton equation and the Kepler laws, or  the Maxwell equations
or  the Schrodinger and Dirac equations are in  reasonable
agreement with experimental data. We can only suspect that these
equations are grounded on the same fundamental principle(s) which
is (are) not known or these equations can be transformed into each
other.

Bohr and Schrodinger assumed that the laws of physics that are
valid in the macrosystem do not hold in the microworld of the
atom. We think that the laws in macro- and microworld are the
same.

We proposed a new mechanism of LENR: cooperative processes in the
whole system - nuclei+atoms+condensed matter - nuclear reactions
in plasma - can occur at smaller threshold energies than the
corresponding ones on free constituents. The cooperative processes
can be induced and enhanced by low energy external fields,
according to the universal resonance synchronization principle.
The excess heat is the emission of internal energy and
transmutations at LENR are the result of redistribution inner
energy of the whole system.

We were able to quantize  phenomenologically (numerology) the
first time the differences between atomic and nuclear rest masses
by the formula (in MeV/$c^{2}$)

$$\Delta M=0.0076294*\frac{n_{1}}{n_{2}},\;
n_{i}=1,2,3,... \eqno(36)$$

Note that this quantization rule is justified for atoms and nuclei
with different $A,\;N$ and $Z$, and the nuclei and atoms represent
a coherent synchronized systems - a complex of coupled oscillators
(resonators). It means that nucleons in nuclei and electrons in
atoms contain all necessary information about the structure of
other nuclei and atoms. This information is used and reproduced by
simple rational relations, according to the fundamental
conservation law of energy-momentum. We originated the universal
cooperative resonance synchronization principle and this principle
is the consequence of the conservation law of energy-momentum. As
a final result the nucleons in nuclei and electrons in atoms have
commensurable frequencies and the differences between those
frequencies are responsible for creation of beating modes. The
phase velocity of standing beating waves can be extremely high;
therefore, all objects of the Universe should get information from
each other almost immediately (instantaneously) using phase
velocity \cite{1,3}. Remember that the beating (modulated) modes
are responsible  for radio and TV-casting.

Therefore, we came to understand the Mach principle.  There are
the different interpretations of the Mach principle. The Mach
principle  can be viewed as an entire Universe being altered by
changes in a single particle and vice versa.

$\otimes$ The universal cooperative resonance synchronization
principle is responsible for the very unity of the Universe.

We have shown only a very small part of our calculations by
formula (36) and the corresponding comparison with experimental
data for atomic and nuclear rest mass differences. This formula
produces a surprisingly high accuracy description of the existing
experimental data. Our noncomplete tentative analysis has shown
that the quantization of rest mass differences demonstrated a very
interesting periodical properties in the whole Mendeleev Table of
Chemical Elements. We hope that it is possible to create an analog
of the Mendeleev Table describing atomic and nuclear properties of
the atomic and nuclear systems simultaneously.

We have proved \cite{1,3} the homology of atom, molecule (in
living molecules too including DNA) and crystal structures. So
interatomic distances in molecules, crystals and solid-state
matter can be written in the following way:

$$ d=\frac{n_{1}}{n_{2}}\lambda_{e}, \eqno(37)$$
where $\lambda_{e}$=0.3324918 nm is the de Broglie electron
wavelength in a hydrogen atom in the ground state
($\lambda_{e}=\lambda_{p}$ in a hydrogen atom in the ground state)
and $n_{1}(n_{2})=1,2,3,...$

In 1953 Schwartz \cite{SCH53} proposed to consider the nuclear and
the corresponding atomic transitions as a united process. This
process contains the $\beta$-decay which represents the transition
of nucleon from  state to state with emission of electron and
antineutrino, and simultaneously the transition of atomic shell
from the initial state to the final one. A complete and strict
solution of this problem is still needed.

We did the first step to consider the nuclear and atomic rest
masses as a unified processes (coupled resonators) which led us to
establish the corresponding phenomenological quantization formulas
(35) and (36), and can bring new possibilities for inducing and
controlling nuclear reactions by atomic processes and new
interpretation of self-organizations of the hierarchial systems in
the Universe including the living cells.

LENR can be stimulated and controlled by the superlow energy
external fields. If frequencies of external field are
commensurable with frequencies of nucleon and electron motions
than we should have resonance enhancement of LENR.  Anomalies LENR
in condensed matter (in plasma) and many anomalies in different
branches in science and technologies ( for example, homoeopathy,
influence of music in nature, nanostructures…) should be results
of  cooperative resonance synchronization frequencies of
subsystems with open system frequencies, with surrounding and
external field frequencies.  In these cases a threshold energy can
be drastically decreased by internal energy of the whole system -
the systems are going to change their structures  if a more stable
systems result. Therefore we have now real possibilities to
stimulate and control many anomalies phenomena including LENR.

\section{Warning}

For nuclear physicists, the LENR (cold fusion) contradicts a
majority of what they have learned throughout most of their
professional lives \cite{KRI05}. The highly specialized profession
created a narrow departmental approach to prediction of global
catastrophes. The ignorance of the whole physical society of LENR
and the lack of financial support may lead to  catastrophes:

$\otimes$The mechanism of shortening the runaway of the reactor at
the Chernobyl Nuclear Power Plant and catastrophes induced by the
HAARP (High Frequency Active Auroral Research Program) program or
by other human activities  may be based on our postulated
cooperative resonance synchronization mechanism.

$\otimes$The same mechanism should be responsible for the ITER
(The International Thermonuclear Experimental  Reactor) explosion
in future.

$\otimes$A. Lipson and G. Miley (ICCC12) showed that the walls of
the ITER TOKAMAK could be damaged by low energy $d$, $t$, $He$ and
intense soft $X$-ray quanta.

$\otimes$The storage of the nuclear waste in stainless steel
containers is a source of real explosion in future. Such explosion
can be caused by radiations of HAARP, for example.

$\otimes$The attack on the World Trade Center (WTC) by terrorists
was a trigger for the low energy nuclear reactions. The whole
destruction of the WTC is a result of the LENR which is  very easy
to prove by isotopic  analysis of stainless steel in towers.

\end{document}